\documentclass[%
 reprint,
 amsmath,amssymb,
 aps,
]{revtex4-2}

\usepackage{graphicx}
\usepackage{dcolumn}
\usepackage{bm}
\usepackage{ulem}[normalem]
\usepackage{xcolor}
\usepackage{float}
\usepackage{placeins}

\begin{document}

\title{High-$z$ SMBHs in Cosmological Models with Enhanced Power Spectra} 
\author{M.V. Tkachev}\email{mtkachev@asc.rssi.ru}
\author{S.V. Pilipenko}%
 \email{spilipenko@asc.rssi.ru}
\author{E.V. Mikheeva}
\email{helen@asc.rssi.ru}
\author{V.N. Lukash}
\email{lukash@asc.rssi.ru}
\affiliation{%
Astro Space Center of P.N. Lebedev Physical Institute, Moscow, Russia}



\date{\today}

\begin{abstract}
We consider the impact of non-power-law spectra of matter perturbations with a bump or with a blue tilt at small scales on the evolution of supermassive black holes (SMBHs) located the innermost part of galaxies. We study SMBH's mass growth rate and the epochs of their birth in four cosmological models using N-body simulations of dark matter within the box of $(100$~Mpc$/h)^3$. The simulations were populated with SMBHs using TRINITY semi-analytic model. We found that the most massive SMBHs at the redshifts $z=5-6$ are similar in all considered cosmologies, including the standard $\Lambda$CDM model. At larger $z$ the bumpy spectra can provide a tenfold increase (relative to $\Lambda$CDM model) in the masses of individual black holes without requiring super-Eddington accretion or heavy seeds. The nucleation of SMBHs starts at $z\sim 13$, which is earlier than in the standard $\Lambda$CMD model.
\end{abstract}

\maketitle


\section{\label{introduction}Introduction}

The James Webb Space Telescope (JWST) has revolutionized the study of high-redshift active galactic nuclei and the early coevolution of supermassive black holes (SMBHs) with their host galaxies. It has revealed a new population of high-redshift ($z>5$), low-luminosity active galactic nuclei in deep surveys and detected the host galaxy's stellar light in the most luminous, massive quasars at $z > 7$, and high number of massive galaxies at $z\geq 9$ \citep{Naidu2022b, Castellano22, Finkelstein22, Donnan23, Labbe23}. Recent findings claim that SMBHs in these systems are significantly more massive than predicted by the standard $\Lambda$CDM model and that this is not due to sample selection effects (see \cite{Li_2025_ApJ_981} and references therein). 

The study of the first SMBHs is particularly important for our understanding of the assembly of galaxies in the Universe (see \cite{Akins_2025} and references therein). It is actively debated now, whether these findings are in tension with the abundance of dark matter (DM) haloes predicted by the $\Lambda$CDM \citep{Boylan-Kolchin23, Lovell22, Chen23, Prada23, Shen23}. 

Recently, non-trivial spectra of density perturbations have attracted considerable interest \cite{Hutsi23, inomata, 2024arXiv240214079R, lukash25}. Such spectra are a significant extension of power-law spectrum of density perturbations. 
Contrary to power-law, they are capable of providing an additional enhancement at small scales, which can be followed by the Population III stars \cite{hirano15}, an excess of high-$z$ galaxies \cite{parashari23, padma23, tkachev23, hirano24}, a formation of \textit{compact} low-mass dark matter halos \cite{yura2024}, and, probably, an early appearance of SMBHs in massive halos. If an enhancement of a power spectrum is high, primordial black holes can be born (see \cite{inomata} and the references therein). 

In this paper, we consider two kinds of primordial spectrum of density perturbations. The first one has a Gaussian bump, the second type of spectra is similar to a \textit{blue-tilted spectrum} considered in \cite{hirano15, hirano24}), therefore here we keep this name.    

The simplest way to produce a power-law spectrum with a bump is to consider a single-field inflation with a kink in the potential (see \cite{star, inn}), or a resonance \cite{ZZPeng}. Since the exact shape of the spectrum can vary and depends on the inflationary model parameters, we use a phenomenological approach, considering a single Gaussian bump added to the smooth power-law spectrum. As for the blue-tilted spectrum, it can be obtained in some variants of double inflation (\cite{lm2000a, lm2000b}), in hybrid inflation (for example, \cite{Clesse2015}), in single-field inflation with specific potential \cite{GarciaMoralez2017} (including potential with inflection point \cite{inflec1, inflec2, inflec3, inflec4}), and in other ways (see a detailed review of \cite{inomata}). 
Usually, spectra with features are considered in the context of primordial black holes, but similar features with smaller amplitudes can be realised on larger scales and influence the evolution of dark matter halos. 

These kind of models have been already considered in \cite{tkachev23, yura2024, PRD2024}, where the halos mass function (HMF), their evolution with redshift, an appearance of compact low-massive halos nearby massive halos, and an inner structure of dark matter halos were studied. In this paper, we concentrate on the epoch and evolution of SMBHs located in the innermost part of galaxies.

The coevolution of SMBHs and their host galaxies has become a "hot topic" in astrophysics after the detection of high-$z$ galaxies and quasars by JWST. It seems that a newly formed SMBH ignites a star formation in the most massive galaxies with high accretion rate, feeding both a seed of SMBH and host galaxies \cite{SilkBegelman2024, Harikane2025}.
Objects named \textit{little red dots} (LRD) in the literature are associated with SMBHs with masses of $10^7-10^8$. Their mass appearance at $z\sim 5$ means that this is when the main generation of LRDs appears, but black holes in massive galaxies at $z\gtrsim 10$ have been detected in the last few years.   
Future observations may reveal even earlier SMBHs and provide clearer constraints on their redshift distribution.

At low redshifts, many relationships between SMBH and host galaxies are established. Black hole (BH) masses are well correlated with properties of the host galaxy such as the projected central velocity dispersion \cite{Ferrarese2000, Tremaine2002, VolNat2009}, the stellar mass, or the luminosity of the bulge component \cite{Magorrian1998, Haring2004}. Although obtained indirectly, other investigations also suggest a correlation with the mass of DM haloes associated with the host galaxy \cite{Ferrarese2002, Baes2003}, an indication that black hole growth follows an evolutionary track, similar to hierarchical structure formation \cite{Silk2010}. It seems that $\Lambda$CDM cosmological simulations are unable to form SMBHs with masses above $10^9 M_\odot$ at $z\simeq 6$, whose existence is inferred from observational evidence \cite{Silk2010}.

Simulations indicate that mergers are dominant processes in the growth of DM halos. The ratio between accreted diffuse matter and a mass obtained in mergers is always less than unity in all cosmological models including the models considered in this paper. As to SMBHs, the situation is not clear but appears to be the opposite (diffuse matter accretion prevails over mergers). It might be related with many different origins, including the result that the black hole accretion rate (BHAR) of baryons is much more effective than the accretion rate of DM, and unclear rate of SMBH mergers, which can be tested by dual SMBH observations \cite{Volonteri2009, Krolik2019, DeRosa2019} .      

BHAR and halo mass growth rate, caused by both merging and accretion of dark matter, are consequences of the same process, which is a hierarchical structure formation. As baryons undergo the gravitational influence of DM, the star formation rate (SFR) is related to these processes too. The activity of a black hole can initiate or prevent star formation \cite{SilkBegelman2024}. 

In this paper, we compare the early evolution of SMBHs in cosmological models with different spectra of perturbations on small scales and discuss which model fits better current observational data. According to some estimates, most of the accreting SMBHs at high redhifts may be heavily obscured by dust (see, e.g. \cite{Ni20}). The existence of such SMBHs can be reliably confirmed by future far infrared (FIR) missions such as Millimetron \cite{Millimetron14,Millimetron21}. Therefore, our work can be considered as a prediction of what can be expected in various models if we could count all the SMBHs, which will be possible with FIR space observatories.

The paper is organized as follows. In Section~\ref{description} we describe the considered matter spectra and numerical simulations, as well as methods used to relate masses of dark matter halos and SMBHs. In Section \ref{sec:results} we present the results concerning the evolution of the BH mass. In Section \ref{sec:discussion} we discuss them.

\section{\label{description}Model description}
\subsection{Cosmological models}

To investigate the impact of spectrum modification on the evolution and growth of SMBHs, we used power spectra constructed as the product of the standard $\Lambda$CDM spectrum and a certain transfer function $T(k)$. In case of the Gaussian bump, it remains the same as in our previous work \cite{tkachev23, yura2024, PRD2024}:
\begin{equation}
    T(k) = 1 + A \cdot \exp \left( -\frac{(\log(k)-\log(k_0))^2}{\sigma_k^2} \right), 
    \label{bumps}
\end{equation}
where $k$ is a wavenumber, $A$, $k_0$, and $\sigma_k$ are bump parameters, and we assume $\sigma_k=0.1$. 

\begin{figure}
\includegraphics[width=1.0\linewidth]{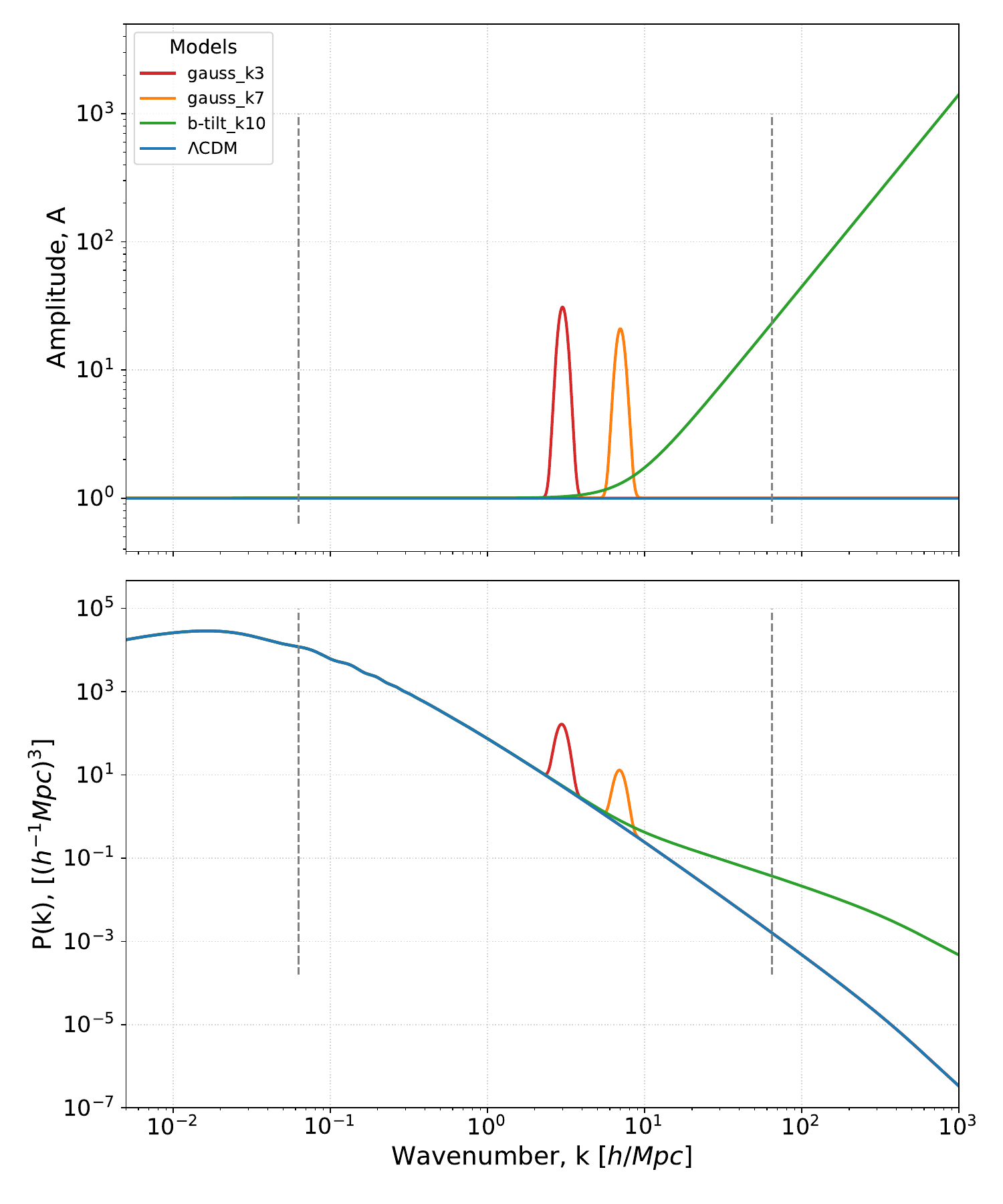}
\caption{The considered spectra of density perturbations (colored lines, see description in text). Grey dashed lines represent the Nyquist frequency for $(5\, $Mpc$/h)^3$ cube and ${2048}^3$ particles resolution. 
\textit{Top panel}: Transfer functions, applied to the $\Lambda$CDM spectrum. 
\textit{Bottom panel:} resulting matter power spectra used in our simulation suite. }
\label{fig:spectra}
\end{figure}

The transfer function for the blue-tilted spectra is calculated as follows:
\begin{equation}
T(k) = \sqrt{1 + \frac{1}{p}\left(\frac{k}{k_0}\right)^{2p+2} },
    \label{tilt}
\end{equation}
where $p$ is a constant. The transfer function essentially defines a smooth transition from $T = 1$ to a power-law function $T(k) = k^{p+1}$, where the parameter $k_0$ defines the value of the wavenumber where the transition occurs. In Table~\ref{tab:sim} we provide the most relevant parameters of the models and simulations considered. Figure~\ref{fig:spectra} illustrates the shapes of the modifications. In both panels, gray dashed lines represent the Nyquist frequency for $(100\, $Mpc$/h)^3$ cube and the resolution of ${2048}^3$ particles. The parameters of each individual spectrum can be found in Table~\ref{tab:sim}.

\begin{table*}
\caption{Most relevant parameters of the simulation suite.} 
\centering
\begin{tabular}{p{0.25\textwidth}p{0.15\textwidth}p{0.15\textwidth}p{0.15\textwidth}p{0.15\textwidth}}
\hline
Main suite:                & \texttt{$\Lambda$CDM}  & \texttt{gauss\_k3}  & \texttt{gauss\_k7}   & \texttt{b-tilt\_k10}   \\ \hline
Box size $($Mpc$/h)$         & 100                  & 100                   & 100                  & 100                  \\
zoomed region resolution   & $2048^3$               & $2048^3$              & $2048^3$             & $2048^3$             \\
Initial redshift           & $300$                  & $1000$                & $1000$               & $1500$               \\
Final redshift             & $5$                    & $5$                   & $5$                  & $5$                  \\
$k_0$                      & --                     & 3                     & 7                    & 10                   \\
$A$                        & --                     & 30                    & 20                   & --                   \\
$p$                        & --                     & --                    & --                   & 0.5                  \\
\hline
\end{tabular}
\label{tab:sim}
\end{table*}

\begin{figure*}
\centering
\includegraphics[width=0.98\textwidth]{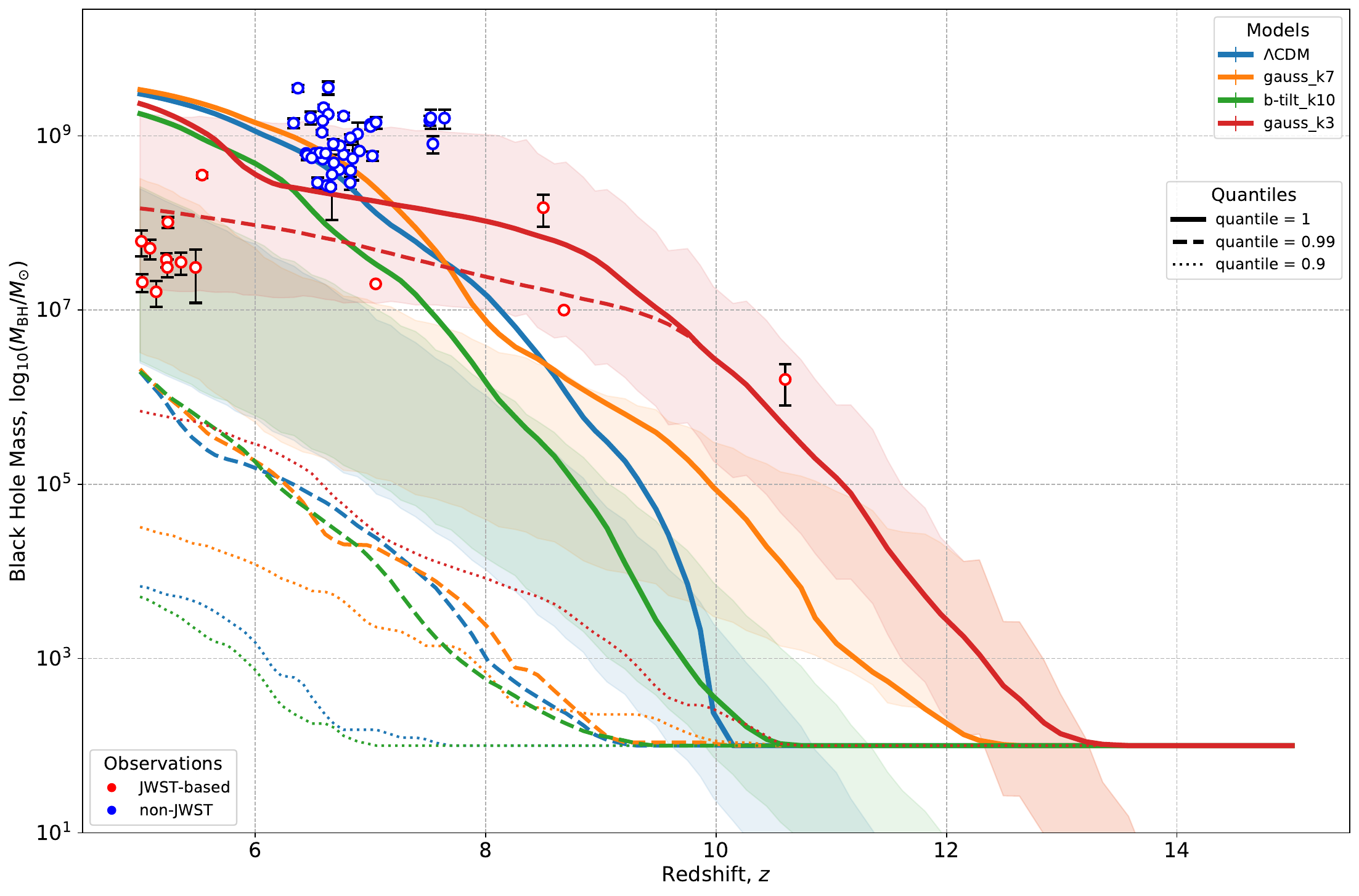}
\caption{Quantile evolution of BH masses in each simulation. The solid, dashed, and dotted lines correspond to the $q=1$, $q=0.99$, and $q=0.9$ mass quantiles at $z=5.0$, respectively. Blue, orange, green, and red lines represent the $\Lambda$CDM, \texttt{gauss\_k7}, \texttt{b-tilt\_k10}, and \texttt{gauss\_k3} simulations. Red circles denote JWST-based observations, blue circles indicate non-JWST detections (see Table~\ref{tab:SMBHs}). The shaded regions represent theoretical mass ranges derived from Press-Schechter (PS) formalism \cite{press}, corresponding to SMBH abundances of 10 (lower boundary) and 0.1 (upper boundary) per simulation volume ($(100~\mathrm{Mpc})^3/160$).}
\label{fig:mass_vs_z}
\end{figure*}

\begin{figure*}
\centering
\includegraphics[width=0.98\textwidth]{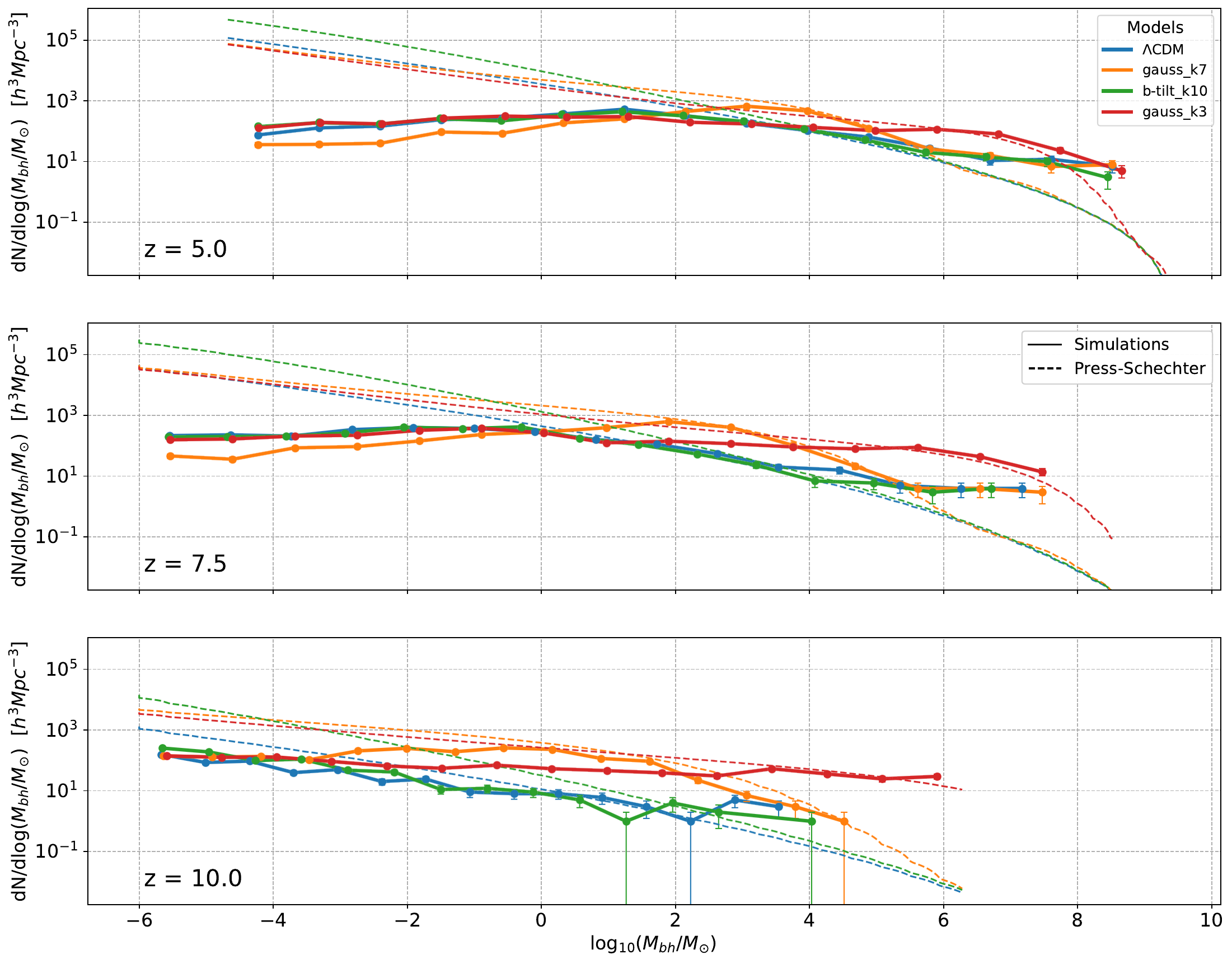}
    \caption{BH mass functions for the four cosmological runs (colored lines) versus a Press-Schechter-based prediction (dashed lines) at three representative redshifts, $z=5$, $z=7.5$, and $z=10$. 
    The vertical axis is $dN/d(\log_{10} M_{\mathrm{BH}})$ per comoving $\mathrm{Mpc}^{-3}$, and the horizontal axis is $\log_{10}(M_{\mathrm{BH}}/M_{\odot})$. 
    Points with error bars show the mass function derived from our simulated BHs, with Poisson uncertainties in each bin. The dashed lines illustrate a Press-Schechter estimate of BH abundance given our assumed seeding and accretion prescriptions. 
    The the abundance of BHs in the intermediate mass range differs more notably for \texttt{gauss\_k7} model, while the \texttt{gauss\_k7} model produces more BHs at the high-end of the mass range.}
    \label{fig:BHAR_hist}
\end{figure*}

\subsection{Numerical simulation}

We have run a series of four dark-matter-only simulations, employing the zoom technique to achieve high resolution in a specific region of interest. Each simulation utilizes a box size of $(100$~Mpc$/h)^3$. Within the zoomed regions (which constitute approximately $1/160$ of the full volume of the cube and are centered around 15 most massive halos), the resolution corresponds to $2048^3$ particles, while intermediate levels are represented by $256^3$, $512^3$, and $1024^3$ particles, respectively. Three of the simulations utilize modified matter power spectra, while one uses the standard $\Lambda$CDM spectrum for comparison.

The simulations were run using the publicly available N-body code \texttt{GADGET-2} \citep{gadget}, which is widely used for cosmological simulations. This code utilizes a combined Tree + Particle Mesh (TreePM) algorithm to calculate gravitational accelerations for each particle by decomposing the gravitational forces into a long-range term and short-range term interaction. Notably, \texttt{GADGET-2} is designed for MPI parallelization, resulting in faster execution and scalability, allowing the code to handle a large number of particles with reasonable computational resources.

To account for the potential early formation of virialized structures, the simulation with tilted spectrum starts at $z = 1500$, the simulations with Gaussian bump spectra start at $z = 1000$, while the $\Lambda$CDM simulation starts at $z = 300$. The final redshift for all simulations is set to $z = 5$. This choice aims to minimize potential artifacts arising from the space periodicity of initial conditions within the simulation box.

The initial conditions for the simulations are generated using the publicly available code \texttt{ginnungagap}\footnote{https://github.com/ginnungagapgroup/ginnungagap}. The matter power spectrum for each simulation is defined individually by applying the appropriate transfer function. For the $\Lambda$CDM simulation, the power spectrum is generated using the publicly available code \texttt{CLASS} \citep{CLASS}. Importantly, the same initial random seed number is used for all simulations, ensuring that they differ only in the amplitude of the power spectrum. Additionally, the amplitude of the longest wavelength mode in the generated initial conditions falls within 20\% of the theoretical value, mitigating the impact of cosmic variance on the high-mass end of the halo mass functions (HMFs).

For each simulation, 100 snapshots are stored at redshift intervals equally spaced in logarithmic scale, ranging from $z = 20$ to $z = 5$. The halo analysis is subsequently performed using the publicly available code \texttt{AHF} \citep{AHF}. This analysis assumes that each halo comprises at least 50 particles, employs a virial overdensity criterion of 200 $\rho_{crit}$, and the spatial resolution of the grid is limited to $100/2^{18}$~Mpc$/h$. 

All simulations share the same cosmological parameters in agreement with the values obtained by \cite{planck}, i.e. $\Omega_m=0.31$, $\Omega_{\Lambda}=1-\Omega_m=0.69$, $\Omega_b=0.048$, $h=0.67$, $n_s=0.96$.

\subsection{From dark matter halo to SMBH
}

Halo masses and redshifts from our simulations are connected to SMBH masses using the \texttt{TRINITY} semi-analytic model \cite{trinity_i, trinity_ii, trinity_iii, trinity_iv}. \texttt{TRINITY} is a flexible empirical framework that statistically connects DM haloes, galaxies, and SMBHs over a broad redshift range ($0 < z < 10$). It is constrained by a wide set of galaxy and AGN observables, including stellar mass functions, SFRs, quasar luminosity functions, and local 
$M_{\bullet}-M_{buldge}$ (black hole–bulge mass) relations. The model infers SMBH accretion rates, merger rates, Eddington ratios, and the evolution of SMBH masses as functions of halo mass, galaxy stellar mass, and redshift. Unlike forward simulations, \texttt{TRINITY} uses population-level statistics rather than individual halo histories, allowing rapid evaluation via Markov Chain Monte Carlo sampling \footnote{The \texttt{TRINITY} code is publicly available at \texttt{https://github.com/HaowenZhang/TRINITY}.}.

In the TRINITY framework it is assumed that all SMBHs are seeded at early cosmic times (with the model effectively initiating SMBH growth from \(z\sim15\)), while no new SMBH seeds are introduced at lower redshifts. This assumption is motivated by the need to maintain a high SMBH occupation fraction in massive haloes down to \(M_{\rm peak} \sim 10^{11}\,M_{\odot}\); without the addition of new seeds, the occupation fraction would otherwise decrease as less massive, unseeded haloes grow in mass \cite{trinity_i}.
Moreover, it is further emphasized that, in order to reconcile the rapid super-Eddington growth inferred at very high redshifts ($z \gtrsim 12$) with the observed properties of active galactic nuclei at $4 \lesssim z \lesssim 11$, the initial seed masses must be relatively low, while the seed properties are implicitly determined by the self-consistent inference of SMBH growth histories.

In practice, for each DM halo we take the mass at a given redshift and match it to the corresponding entry in the \texttt{TRINITY} tables, thereby assigning an SMBH mass and accretion rate to that halo. This procedure yields the SMBH mass and BHAR associated with each halo at each output time, which we then use for further analysis. In particular, \texttt{TRINITY} relies on the HMF at each redshift as an input to infer the halo–SMBH connection.
This procedure results in the appearance of artificial BHs with masses less than $1\,M_\odot$: it provides a necessary number of black holes.

While \texttt{TRINITY} uses $\Lambda$CDM HMF and merger rate to build a connection between the halo mass and BHAR, we have applied it to all our cosmological models, including those with modified primordial power spectra (e.g., the blue-tilted spectrum). Our modified power spectra differ from the $\Lambda$CDM one only at relatively high wavenumbers $k\geq3$, as a result, at later times, when $z\ll 10$, the HMF for the considered spectra only slightly differ from the $\Lambda$CDM one. Since \texttt{TRINITY} uses low-redshift observational data, we assume that the connection between the halo mass and BHAR found in \texttt{TRINITY} is only slightly affected by the change in the power spectrum we consider.


For the majority of our analysis, we assume that all SMBHs begin with a seed mass of $M_{\rm seed} = 100\,M_\odot$, and that their subsequent mass growth assumes Eddington-limited accretion (i.e. an effective accretion efficiency was assumed \(\eta=1\); however, lowering the \(\eta\) even by a factor of 10 would not affect the overall outcome of the models). In this context, the SMBH mass is computed via the integral
\begin{equation}
M_{\rm BH}(t) = M_{\rm seed} + \int_{t_{\rm seed}}^{t} \eta\,\dot M_{\mathrm{BHAR}}(t')\,dt',
\end{equation}
where 
BHAR values $\dot M_{\mathrm{BHAR}}$ are taken from \texttt{TRINITY}. This methodology is applied in most of our analysis to fairly compare the mass growth histories computed from the known BHAR and Eddington accretion rate.

We have also tested the impact of varying the SMBH seed mass, increasing it from $100,M_\odot$ to $1000,M_\odot$. We find that this change does not significantly affect the formation or growth histories of the massive SMBHs in our simulations. This insensitivity arises because the semi-analytic TRINITY model assigns accretion rates and black hole masses to halos based on empirical calibration to observed populations, independently of the adopted seed mass, as long as the seeds are introduced early enough and are much less than the final masses. Thus, our main conclusions regarding the early emergence and abundance of massive SMBHs are robust to the assumed seed mass.

However, when constructing the BH mass function (see Fig.~\ref{fig:BHAR_hist} in Section \ref{sec:results}), we plot the SMBH masses directly computed by the \texttt{TRINITY} model. This is because the integration method, by design, cannot produce BH masses below the seed mass (100\(M_\odot\)); therefore, it would impose a cutoff at the low-mass end. In contrast, the direct \texttt{TRINITY} output includes BH masses below this threshold, and we adopt those values when plotting the mass function to ensure that the low-mass end is fully represented for comparison with Press–Schechter (PS) estimates \cite{press} (here we used the Sheth-Tormen extension \cite{Sheth99}), even if such sub-\(100\,M_\odot\) masses may be considered atypical in conventional models.

\section{Results \label{sec:results}}

\subsection{Black Hole Mass Evolution}
\label{subsec:bh_mass_evolution}

Using \texttt{TRINITY},
we examine the evolution of SMBH masses in our four cosmological simulations. Figure~\ref{fig:mass_vs_z} illustrates BH mass evolution through three BH mass quantiles: $q=1$ (which simply represents the most massive black hole at each snapshot), $q=0.99$ (mass above 99\% of SMBHs), and $q=0.9$ at the moment $z=5$. Note that at higher redshifts the mass quantiles could have been different, e.g. the most massive BH at $z=5$ was not the most massive at $z = 10$.


\begin{table*}[ht]
\caption{Observational data on high-$z$ SMBHs from various sources.}
\centering
\begin{tabular}{p{0.05\textwidth}p{0.50\textwidth}p{0.15\textwidth}p{0.25\textwidth}}
\hline
\textbf{\#} & \textbf{Black Hole Mass} & \textbf{Redshift} & \textbf{Source} \\
\hline
1 & $\log_{10}(M_{\mathrm{BH}}/M_{\odot}) = 8.17 \pm 0.42$ & 8.50 & Kokorev et al.\ (2023) \\

2 & $\sim 2\times 10^{7}\,M_{\odot}$ (broad H$\beta$; no explicit error reported) & 7.045 & Furtak et al.\ (2023) \\

3 & 20 faint AGNs with $M_{\mathrm{BH}}\sim 10^{7-8}\,M_{\odot}$ & 4.2--5.5 & Matthee et al.\ (2024) \\

4 & $1.5\times 10^{9}\,M_{\odot}$ (quoted $\pm0.2\times10^9$) & 7.515 & Yang et al.\ (2020) \\

5 & $(1.6\pm 0.4)\times 10^{9}\,M_{\odot}$ & 7.642 & Wang et al.\ (2021) \\

6 & $\sim 10^{7}\,M_{\odot}$ (CEERS\_1019) & 8.679 & Larson et al.\ (2023) \\

7 & $\sim 1.6\times 10^{6}\,M_{\odot}$ (unc.\ $\pm0.3$\,dex) & 10.60 & Maiolino et al.\ (2024) \\

8 & $\sim 10^{6}\,M_{\odot}$ (CEERS-AGN-z5-1) & 5.0 & Onoue et al.\ (2023) \\

9 & 37 quasars with $M_{\mathrm{BH}}=(0.3\text{--}3.6)\times 10^{9}\,M_{\odot}$ & 6.3--7.64 & Yang et al.\ (2021) \\
\hline
\end{tabular}
\label{tab:SMBHs}
\end{table*}

All four models possess at least some BHs that undergo early and rapid growth, producing roughly the same $q=1$ curve by $z\sim 6-7$. In contrast, even for the $q=0.99$ quantile the BH mass in all models, with the exception of \texttt{gauss\_k7}, differs by at least 2 orders from the $q=1$ case. The difference becomes more noticeable at higher redshifts, where enhanced small-scale power significantly boosts the population of moderately massive SMBHs. This leads to a clear separation in the 99\% quantile at $z \gtrsim 8$. The \texttt{gauss\_k7} model shows a similar, albeit less pronounced, early mass buildup. By $z \approx 5$, differences between the models decreases as enhanced structure growth diminishes. 

The shaded regions in Figure~\ref{fig:mass_vs_z}, derived from a PS formalism, provide a theoretical reference for the likelihood of encountering a BH of a given mass within a smaller sub-volume. Specifically, the lower boundary marks the BH mass above which a region of $(100\,\mathrm{Mpc})^3/160$ would be expected to include approximately 10 such SMBHs, while the upper boundary corresponds to a 10\% probability of finding a BH of that mass in the same sub-volume. These PS bounds are centered on the extremes of the mass distribution in a limited volume, providing context for the results of our simulation. Size of our simulation cube, $(100$~Mpc$/h)^3$, is substantially smaller than the Hubble volume, therefore, the halos hosting SMBHs are expected to be considerably more common than what would be inferred from a full-volume, homogeneous estimate.


The overplotted data points in Figure~\ref{fig:mass_vs_z} are observational data points for high-redshift SMBHs (see Table \ref{tab:SMBHs} for details). We depict the observations from JWST-based surveys in red circles and non-JWST observations in blue circles. It is evident from the figure that our $\Lambda$CDM run is not capable of producing the $\sim10^9 M_\odot$ quasars seen at $z\sim7.5$ (entries 4–5 in Table~2) within our limited volume. On the other hand, the \texttt{gauss\_k3} simulation, featuring the largest enhancement of small-scale power, easily produces massive SMBHs at the highest redshifts ($z \sim 8-10$). This model easily accommodates even the most extreme observational data points (particularly \cite{maiolino2024} at $z=10.6$), suggesting that such SMBHs may not represent rare or exceptional occurrences within cosmologies characterized by enhanced primordial power. 

Figure~\ref{fig:BHAR_hist} compares the BH mass functions from our four models to a PS calculation at three illustrative redshifts: $z=5$, $z=7.5$, and $z=10$. Here BH masses for the mass function were taken from \texttt{TRINITY} directly, contrary to the Fig.~\ref{fig:mass_vs_z}, where we calculated the resulting BH masses from BHAR. 
The simulations are shown as points with Poissonian error bars, while the dashed lines are the PS-based analytical estimates. 
At $z=5$, the highest BH masses ($\log_{10} M_{\mathrm{BH}} \gtrsim 6$) behave similarly across all runs, consistent with the $q=1$ curve in Figure~\ref{fig:mass_vs_z} and confirming that the emergence of the most massive BHs at lower redshifts is largely independent of the modifications to small-scale power.
However, at higher redshifts the \texttt{gauss\_k3} simulation produces a notably greater abundance of high- and intermediate-mass BHs ($M_{\mathrm{BH}}\sim10^{5}-10^7\,M_{\odot}$), compared to the other models. The \texttt{gauss\_k7} run shows a significantly stronger abundance of low-mass BHs at $z\gtrsim 7$ ($M_{\mathrm{BH}}\sim10^{2}-10^3\,M_{\odot}$).
By $z \approx 5$, the four scenarios converge closer together, mirroring what is seen in the quantile plot:
once a large fraction of halos have collapsed and begun feeding their BH seeds, the initial differences in the small-scale structure become less critical.  

 
In summary, our results demonstrate that all four models are capable of producing rare, massive SMBHs by $z\sim 6-7$, although the typical high-end population (e.g. the 99th percentile) shows a marked dependence on the underlying power spectrum, with models exhibiting enhanced small-scale power generating a more abundant population of moderately massive BHs at high redshifts. Moreover, despite the limited volume of our simulation cube relative to the full cosmological volume, the Press–Schechter reference indicates that the halo abundance in our simulation is sufficient to account for the observed SMBH distribution without requiring extreme rarity. These findings, taken together, suggest that within cosmologies featuring enhanced primordial power, the early formation of massive BHs may be more efficient and less exceptional than previously assumed.

\subsection{Black Hole Accretion Rate}

\begin{figure*}
\centering
\includegraphics[width=0.98\textwidth]{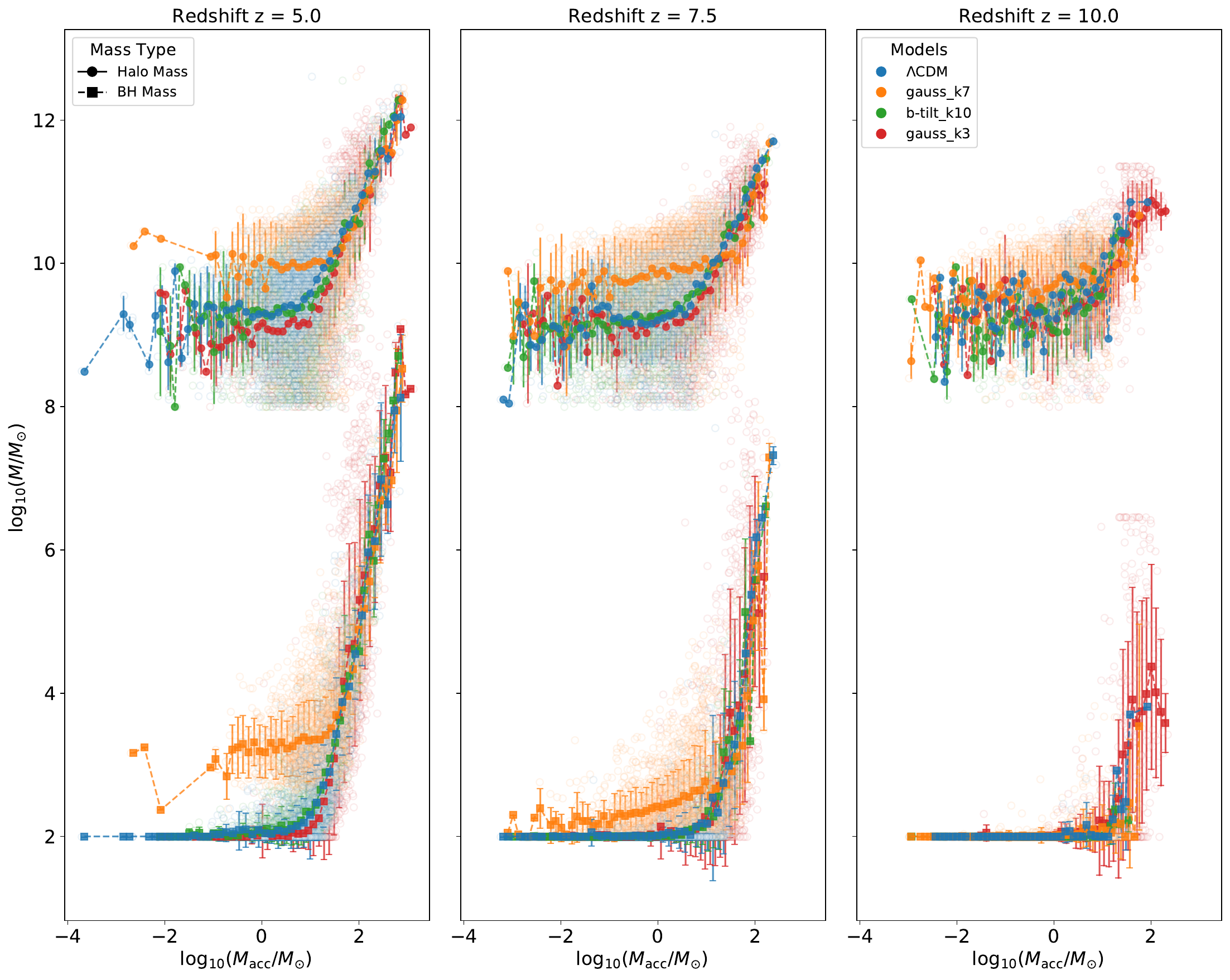}
    \caption{Relationship between halo mass (solid lines, circles) and BH mass (dashed lines, squares) as a function of the merger-induced mass accretion rate. Different colors represent distinct cosmological simulations. Each data point corresponds to a binned average, and error bars indicate the standard deviation within each accretion rate bin.}
    \label{fig:M_vs_Macc}
\end{figure*}
To clarify the impact of matter spectrum on BHAR, 
halo merger-driven accretion rates were computed by tracking halo particles across simulation snapshots. For each halo at a given snapshot, we identified all DM particles that were not part of its main progenitor in the previous snapshot. This was accomplished by tracing the halo particle IDs backward in time and excluding those already belonging to the primary progenitor halo. Only particles coming from secondary progenitors are counted as newly accreted via mergers. Then we calculate the fraction of such particles and multiply it by the total mass of the halo at given snapshot, which yields the mass gained through mergers; then divide it by the cosmological time interval between the snapshots, which gives the merger-induced accretion rate $\dot{M}_{\rm merger}$.

Figure~\ref{fig:M_vs_Macc} shows that both the halo mass $M_{\rm halo}$ and BH mass $M_{\rm BH}$ scale positively with this merger-driven accretion rate. Halos experiencing higher $\dot{M}_{\rm merger}$ tend to be more massive, and their central SMBHs are correspondingly more massive. Both $M_{\rm halo}$ and $M_{\rm BH}$ exhibit an approximately power-law dependence on $\dot{M}_{\rm merger}$. Notably, the $M_{\rm BH}$ versus $\dot{M}_{\rm merger}$ relation is tighter (with less scatter) than the $M_{\rm halo}$ versus $\dot{M}_{\rm merger}$ relation, indicating a stronger correlation between BH growth and merger-driven halo growth.

This tighter coupling suggests that SMBH growth is highly episodic and closely tied to discrete halo merger events, whereas halo mass assembly includes substantial continuous, "smooth" accretion that is not captured by the merger-only rate. In other words, a significant portion of a BH’s mass is accumulated during bursty accretion episodes triggered by mergers, while halos continue to grow between mergers via diffuse accretion, leading to more scatter in the halo trend. Such a power-law co-evolution is supported by theoretical frameworks like the \texttt{TRINITY} empirical model\cite{trinity_i, trinity_ii, trinity_iii, trinity_iv}, which links BH accretion to host halo assembly and finds that SMBHs co-grow with their halos in tandem. This implies that the observed correlation is consistent with a scenario in which BHs preferentially gain mass during merger-induced accretion periods, reinforcing the idea of synchronized, merger-driven growth for halos and their central SMBHs.

Comparing panels of Figure~\ref{fig:M_vs_Macc} we see that the model \texttt{gauss\_k3} demonstrates the tendency to a earlier appearance of massive BH in a central part of halos, whereas the halo masses are similar in all models. Also, it is interesting to remark that at fixed $\log(M_{acc}/M_\odot)$ the highest values of both halo and BH masses are achieved with \texttt{gauss\_k7} model. 

\section{Discussion \label{sec:discussion}}
While our model of SMBH growth under non-standard primordial power spectra yields intriguing results, it also comes with 
some 
limitations. We discuss these here to clarify the scope and reliability of our findings. 

\paragraph{Validity of \textsc{TRINITY} for nonstandard cosmologies.}
A key approximation in our modeling is the use of \texttt{TRINITY}, a semi-analytic framework originally calibrated within the standard $\Lambda$CDM cosmology, to compute BHAR for halos in all four of our simulations -- including those with modified initial power spectra. Specifically, \texttt{TRINITY} uses the HMF at each redshift as an input to derive the statistical mapping from halo mass to SMBH accretion rate and related quantities. In our analysis, we input the $\Lambda$CDM HMF into \texttt{TRINITY} in all models. This effectively assumes that the functional form of $\mathrm{BHAR}(M_h, z)$ remains the same in each cosmology. 
However, the halo masses themselves come directly from our simulations and differ across models. Therefore, even though the $\mathrm{BHAR}(M_h, z)$ relation is shared for all models, the resulting SMBH growth histories are quite distinct, driven by the different halo formation rates in each scenario.

As a positive argument to use the $\Lambda$CDM relation is the universality of density profiles in non-trivial cosmologies, found and studied in \cite{PRD2024}. Also, as has been shown in \cite{tkachev23}, the low-$z$ HMF for different power spectra considered here approaches to the $\Lambda$CDM one, so the impact of power spectrum on \texttt{TRINITY} calibration should be moderate, since it uses low-$z$ data. Truly self-consistent model would require recalibrating this mapping for each cosmology, likely through hydrodynamical simulations or forward modeling of observable galaxy and active galaxy nuclei populations. 

\paragraph{SMBH mergers treated via halo growth.} Another caveat is that our model does not explicitly track SMBH mergers. The \textsc{TRINITY} approach 
focuses on reproducing observed SMBH scaling relations and growth rates, primarily through continuous accretion, fueled by halo and galaxy growth. Thus, when two halos (each possibly harboring a central SMBH) merge in our framework, we do not simulate the dynamical inspiral and coalescence of the two black holes. Instead, the effect of the merger is implicitly explained by allowing the new larger halo (and its central galaxy) to host a correspondingly more massive SMBH, determined by the empirical halo–galaxy–SMBH relations. In other words, we assume that the SMBH in the primary halo effectively absorbs the mass of the incoming SMBH, but this is done via a    growth, rather than an actual binary BH merger event. This simplification is common in semi-analytic models, but it is only an approximation to reality. Actual SMBH mergers can involve complexities like gravitational radiation (which can carry away a fraction of the mass and energy) and delays due to dynamical friction and hardening of the BH binary. Our model does not capture those details; it simply adds mass as halos grow, without enforcing strict mass conservation for individual SMBH merger events. As a result, there is some uncertainty in our SMBH mass budgets -- for example, we might slightly overestimate the final mass if we implicitly double-count mass that should have been lost to gravitational waves, or underestimate it if a merged secondary SMBH would have continued accreting on its own for some time. We expect these effects to be second-order for global statistics, but they nonetheless represent a limitation of our modeling. A more rigorous treatment would require following SMBH pair dynamics and coalescences explicitly, which lies beyond the scope of our current semi-analytical approach. 

\paragraph{Bias from zoom-in simulation regions.} A further limitation comes from the nature of our simulations, which were zoom-in simulations centered on high-density peaks in the early Universe. By design, these simulations target rare, overdense regions to efficiently study the formation of very massive SMBHs. The drawback is that these regions are not representative of the average patch of the universe. They constitute a biased subset of volume where structure formation is accelerated and amplified. Future work could address this by running a larger suite of simulations covering a range of environments or by applying weighting schemes to statistically recover global averages, but such measures were beyond the scope of this paper.

\paragraph{Neglect of baryonic physics in simulations.} Additionally we would like to emphasize again that our simulations were dark-matter-only, neglecting gas dynamics and other baryonic processes. Baryons (gas, stars, and their associated physics) can substantially alter the formation and growth of both halos and SMBHs. For instance, without gas physics, our simulation cannot model cooling and inflows that supply fuel to a growing SMBH, nor can it capture feedback mechanisms (stellar winds, supernovae, or AGN feedback) that can heat or expel gas and thereby regulate growth.

In our case, because we used \textsc{TRINITY} (which is empirically constrained by real observations of galaxies and AGN), some baryonic effects enter indirectly-\textsc{TRINITY}’s parameters for SMBH growth and radiative efficiency are informed by observed galaxy–SMBH relations, which themselves bear the imprint of cooling and feedback processes. However, this is not a substitute for explicit hydrodynamics. In reality, we expect some of our model predictions (for example, the maximum SMBH masses or the speed of early growth) to be moderated once processes like feedback are accounted for \cite{benson2020}. A more complete analysis would require running full hydrodynamic simulations, which we plan to explore in our future work.

We have attempted to mitigate or at least quantify each of these caveats where possible (e.g. fixing the HMF, comparing to analytic theory, and basing our SMBH growth parameters on real data), but a degree of caution is warranted in extrapolating the results. None of these limitations invalidates our qualitative findings -- indeed, the trends we identified are robust within the context of our models -- but they do suggest that the absolute values and rapidity of SMBH growth in our scenarios should be interpreted with care. Future work, employing more flexible modeling tools and more representative simulations, will be valuable to confirm and extend the results presented here.


\section{Conclusions}

We have performed a set of high-resolution N-body zoom-in simulations to investigate the formation of supermassive black holes (SMBHs) in cosmologies with non-power-law primordial spectra. The focus was on quantifying the impact of enhanced small-scale power on early SMBH seeding and growth, with particular attention to the redshift distribution and mass assembly of the most massive black holes. We considered three non-standard scenarios: two models with a Gaussian bump in the initial power spectrum at $k=3$ and $k=7$~Mpc${}^{-1}$ (\texttt{gauss\_k3} and \texttt{gauss\_k7}), and one model with a scale-invariant blue tilt on small scales (\texttt{b-tilt\_k10}). A standard $\Lambda$CDM simulation was run in parallel for comparison.

To estimate SMBH growth in each cosmology, we used the semi-analytic model \texttt{TRINITY}, which maps dark matter halo mass and redshift to black hole accretion rates. Importantly, the mapping $\mathrm{BHAR}(M_h, z)$ used in \texttt{TRINITY} is calibrated within $\Lambda$CDM and was held fixed across all simulations. Thus, differences in SMBH evolution arise entirely from variations in the halo formation histories caused by the modified initial power spectra. In practice, this means the relationship between black holes and their host halos was assumed universal, while the halo populations themselves were drawn from self-consistent cosmological simulations for each model.

We find that the most massive SMBHs at $z \sim 5–7$ can form across all four cosmologies, including $\Lambda$CDM. However, at earlier times ($z \gtrsim 8$), the differences become striking. In the \texttt{gauss\_k3} model, which features the largest small-scale power excess, the upper quantile of SMBH masses reaches values above $10^9\,M_\odot$ by $z\sim10$, without invoking super-Eddington accretion or exotic seed mechanisms. These early-forming BHs match or exceed the masses of several high-redshift SMBHs observed by JWST and other surveys. In contrast, the $\Lambda$CDM model does not produce a comparable SMBH population at such early epochs. Even more moderate modifications, as in the \texttt{gauss\_k7} run, shift the onset of BH seeding to higher redshifts and increase the abundance of intermediate-mass black holes by $z\sim8$.

In addition, we examined the relationship between SMBH mass and the cumulative black hole accretion rate derived from TRINITY. We find that the black hole mass correlates more strongly with the cumulative merger-driven accretion than with the instantaneous host halo mass (see Figure~\ref{fig:M_vs_Macc}). This supports the interpretation that SMBH growth is episodic and closely linked to discrete merger events. While halo mass growth includes a substantial contribution from continuous accretion of dark matter onto halo, this component appears to play a less direct role in fueling black holes, at least in the framework employed here.

Our results suggest that relatively modest changes to the primordial power spectrum can have a substantial impact on the SMBH population at early cosmic times. The formation of massive black holes at $z > 8$ does not require extreme assumptions, but can arise naturally in cosmologies with slightly enhanced small-scale structure. This provides a promising explanation for the observed population of bright high-redshift quasars. However, several limitations remain. Our modeling assumes a fixed empirical BHAR relation across all cosmologies, neglects baryonic physics, and does not include explicit black hole mergers. Future work will address these caveats using hydrodynamical simulations and recalibrated semi-analytic models to more accurately capture feedback, cooling, and accretion in modified early universes.

Current observational data seem to support the models with enhanced power at small scales, especially the \texttt{gauss\_k3} model. However, existing observatories can see only the tip of the iceberg, with $\sim 80-90$\% of active SMBHs being obscured by dust and gas. With the development of new facilities in the FIR range, such as Millimetron, key diagnostic lines of AGN could be observed even through enormously dense clouds, which will allow to test both the SMBH growth models and the small scale power in the spectrum of perturbations.

\begin{acknowledgments}
Authors are grateful to S.~Drozdov for helpful discussions and the anonymous referee for their valuable remarks.
\end{acknowledgments}


\newcommand{\mnras}{MNRAS}
\newcommand{\jcap}{J. Cosmology Astropart. Phys.}
\newcommand{\apjs}{ApJS} 
\newcommand{\aap}{A\&A} 
\newcommand{\apjl}{ApJ Letters}

\bibliography{refs}

\begin{thebibliography}{64}%
\makeatletter
\providecommand \@ifxundefined [1]{%
 \@ifx{#1\undefined}
}%
\providecommand \@ifnum [1]{%
 \ifnum #1\expandafter \@firstoftwo
 \else \expandafter \@secondoftwo
 \fi
}%
\providecommand \@ifx [1]{%
 \ifx #1\expandafter \@firstoftwo
 \else \expandafter \@secondoftwo
 \fi
}%
\providecommand \natexlab [1]{#1}%
\providecommand \enquote  [1]{``#1''}%
\providecommand \bibnamefont  [1]{#1}%
\providecommand \bibfnamefont [1]{#1}%
\providecommand \citenamefont [1]{#1}%
\providecommand \href@noop [0]{\@secondoftwo}%
\providecommand \href [0]{\begingroup \@sanitize@url \@href}%
\providecommand \@href[1]{\@@startlink{#1}\@@href}%
\providecommand \@@href[1]{\endgroup#1\@@endlink}%
\providecommand \@sanitize@url [0]{\catcode `\\12\catcode `\$12\catcode `\&12\catcode `\#12\catcode `\^12\catcode `\_12\catcode `\%12\relax}%
\providecommand \@@startlink[1]{}%
\providecommand \@@endlink[0]{}%
\providecommand \url  [0]{\begingroup\@sanitize@url \@url }%
\providecommand \@url [1]{\endgroup\@href {#1}{\urlprefix }}%
\providecommand \urlprefix  [0]{URL }%
\providecommand \Eprint [0]{\href }%
\providecommand \doibase [0]{https://doi.org/}%
\providecommand \selectlanguage [0]{\@gobble}%
\providecommand \bibinfo  [0]{\@secondoftwo}%
\providecommand \bibfield  [0]{\@secondoftwo}%
\providecommand \translation [1]{[#1]}%
\providecommand \BibitemOpen [0]{}%
\providecommand \bibitemStop [0]{}%
\providecommand \bibitemNoStop [0]{.\EOS\space}%
\providecommand \EOS [0]{\spacefactor3000\relax}%
\providecommand \BibitemShut  [1]{\csname bibitem#1\endcsname}%
\let\auto@bib@innerbib\@empty
\bibitem [{\citenamefont {{Naidu}}\ \emph {et~al.}(2022)\citenamefont {{Naidu}}, \citenamefont {{Oesch}}, \citenamefont {{van Dokkum}}, \citenamefont {{Nelson}}, \citenamefont {{Suess}}, \citenamefont {{Brammer}}, \citenamefont {{Whitaker}}, \citenamefont {{Illingworth}}, \citenamefont {{Bouwens}}, \citenamefont {{Tacchella}}, \citenamefont {{Matthee}}, \citenamefont {{Allen}}, \citenamefont {{Bezanson}}, \citenamefont {{Conroy}}, \citenamefont {{Labbe}}, \citenamefont {{Leja}}, \citenamefont {{Leonova}}, \citenamefont {{Magee}}, \citenamefont {{Price}}, \citenamefont {{Setton}}, \citenamefont {{Strait}}, \citenamefont {{Stefanon}}, \citenamefont {{Toft}}, \citenamefont {{Weaver}},\ and\ \citenamefont {{Weibel}}}]{Naidu2022b}%
  \BibitemOpen
  \bibfield  {author} {\bibinfo {author} {\bibfnamefont {R.~P.}\ \bibnamefont {{Naidu}}}, \bibinfo {author} {\bibfnamefont {P.~A.}\ \bibnamefont {{Oesch}}}, \bibinfo {author} {\bibfnamefont {P.}~\bibnamefont {{van Dokkum}}}, \bibinfo {author} {\bibfnamefont {E.~J.}\ \bibnamefont {{Nelson}}}, \bibinfo {author} {\bibfnamefont {K.~A.}\ \bibnamefont {{Suess}}}, \bibinfo {author} {\bibfnamefont {G.}~\bibnamefont {{Brammer}}}, \bibinfo {author} {\bibfnamefont {K.~E.}\ \bibnamefont {{Whitaker}}}, \bibinfo {author} {\bibfnamefont {G.}~\bibnamefont {{Illingworth}}}, \bibinfo {author} {\bibfnamefont {R.}~\bibnamefont {{Bouwens}}}, \bibinfo {author} {\bibfnamefont {S.}~\bibnamefont {{Tacchella}}}, \bibinfo {author} {\bibfnamefont {J.}~\bibnamefont {{Matthee}}}, \bibinfo {author} {\bibfnamefont {N.}~\bibnamefont {{Allen}}}, \bibinfo {author} {\bibfnamefont {R.}~\bibnamefont {{Bezanson}}}, \bibinfo {author} {\bibfnamefont {C.}~\bibnamefont {{Conroy}}}, \bibinfo {author} {\bibfnamefont {I.}~\bibnamefont {{Labbe}}},
  \bibinfo {author} {\bibfnamefont {J.}~\bibnamefont {{Leja}}}, \bibinfo {author} {\bibfnamefont {E.}~\bibnamefont {{Leonova}}}, \bibinfo {author} {\bibfnamefont {D.}~\bibnamefont {{Magee}}}, \bibinfo {author} {\bibfnamefont {S.~H.}\ \bibnamefont {{Price}}}, \bibinfo {author} {\bibfnamefont {D.~J.}\ \bibnamefont {{Setton}}}, \bibinfo {author} {\bibfnamefont {V.}~\bibnamefont {{Strait}}}, \bibinfo {author} {\bibfnamefont {M.}~\bibnamefont {{Stefanon}}}, \bibinfo {author} {\bibfnamefont {S.}~\bibnamefont {{Toft}}}, \bibinfo {author} {\bibfnamefont {J.~R.}\ \bibnamefont {{Weaver}}},\ and\ \bibinfo {author} {\bibfnamefont {A.}~\bibnamefont {{Weibel}}},\ }\bibfield  {title} {\bibinfo {title} {{Two Remarkably Luminous Galaxy Candidates at z {\ensuremath{\approx}} 10-12 Revealed by JWST}},\ }\href {https://doi.org/10.3847/2041-8213/ac9b22} {\bibfield  {journal} {\bibinfo  {journal} {\apjl}\ }\textbf {\bibinfo {volume} {940}},\ \bibinfo {eid} {L14} (\bibinfo {year} {2022})},\ \Eprint
  {https://arxiv.org/abs/2207.09434} {arXiv:2207.09434 [astro-ph.GA]} \BibitemShut {NoStop}%
\bibitem [{\citenamefont {{Castellano}}\ \emph {et~al.}(2022)\citenamefont {{Castellano}}, \citenamefont {{Fontana}}, \citenamefont {{Treu}}, \citenamefont {{Santini}}, \citenamefont {{Merlin}}, \citenamefont {{Leethochawalit}}, \citenamefont {{Trenti}}, \citenamefont {{Vanzella}}, \citenamefont {{Mestric}}, \citenamefont {{Bonchi}}, \citenamefont {{Belfiori}}, \citenamefont {{Nonino}}, \citenamefont {{Paris}}, \citenamefont {{Polenta}}, \citenamefont {{Roberts-Borsani}}, \citenamefont {{Boyett}}, \citenamefont {{Brada{\v{c}}}}, \citenamefont {{Calabr{\`o}}}, \citenamefont {{Glazebrook}}, \citenamefont {{Grillo}}, \citenamefont {{Mascia}}, \citenamefont {{Mason}}, \citenamefont {{Mercurio}}, \citenamefont {{Morishita}}, \citenamefont {{Nanayakkara}}, \citenamefont {{Pentericci}}, \citenamefont {{Rosati}}, \citenamefont {{Vulcani}}, \citenamefont {{Wang}},\ and\ \citenamefont {{Yang}}}]{Castellano22}%
  \BibitemOpen
  \bibfield  {author} {\bibinfo {author} {\bibfnamefont {M.}~\bibnamefont {{Castellano}}}, \bibinfo {author} {\bibfnamefont {A.}~\bibnamefont {{Fontana}}}, \bibinfo {author} {\bibfnamefont {T.}~\bibnamefont {{Treu}}}, \bibinfo {author} {\bibfnamefont {P.}~\bibnamefont {{Santini}}}, \bibinfo {author} {\bibfnamefont {E.}~\bibnamefont {{Merlin}}}, \bibinfo {author} {\bibfnamefont {N.}~\bibnamefont {{Leethochawalit}}}, \bibinfo {author} {\bibfnamefont {M.}~\bibnamefont {{Trenti}}}, \bibinfo {author} {\bibfnamefont {E.}~\bibnamefont {{Vanzella}}}, \bibinfo {author} {\bibfnamefont {U.}~\bibnamefont {{Mestric}}}, \bibinfo {author} {\bibfnamefont {A.}~\bibnamefont {{Bonchi}}}, \bibinfo {author} {\bibfnamefont {D.}~\bibnamefont {{Belfiori}}}, \bibinfo {author} {\bibfnamefont {M.}~\bibnamefont {{Nonino}}}, \bibinfo {author} {\bibfnamefont {D.}~\bibnamefont {{Paris}}}, \bibinfo {author} {\bibfnamefont {G.}~\bibnamefont {{Polenta}}}, \bibinfo {author} {\bibfnamefont {G.}~\bibnamefont {{Roberts-Borsani}}}, \bibinfo
  {author} {\bibfnamefont {K.}~\bibnamefont {{Boyett}}}, \bibinfo {author} {\bibfnamefont {M.}~\bibnamefont {{Brada{\v{c}}}}}, \bibinfo {author} {\bibfnamefont {A.}~\bibnamefont {{Calabr{\`o}}}}, \bibinfo {author} {\bibfnamefont {K.}~\bibnamefont {{Glazebrook}}}, \bibinfo {author} {\bibfnamefont {C.}~\bibnamefont {{Grillo}}}, \bibinfo {author} {\bibfnamefont {S.}~\bibnamefont {{Mascia}}}, \bibinfo {author} {\bibfnamefont {C.}~\bibnamefont {{Mason}}}, \bibinfo {author} {\bibfnamefont {A.}~\bibnamefont {{Mercurio}}}, \bibinfo {author} {\bibfnamefont {T.}~\bibnamefont {{Morishita}}}, \bibinfo {author} {\bibfnamefont {T.}~\bibnamefont {{Nanayakkara}}}, \bibinfo {author} {\bibfnamefont {L.}~\bibnamefont {{Pentericci}}}, \bibinfo {author} {\bibfnamefont {P.}~\bibnamefont {{Rosati}}}, \bibinfo {author} {\bibfnamefont {B.}~\bibnamefont {{Vulcani}}}, \bibinfo {author} {\bibfnamefont {X.}~\bibnamefont {{Wang}}},\ and\ \bibinfo {author} {\bibfnamefont {L.}~\bibnamefont {{Yang}}},\ }\bibfield  {title} {\bibinfo {title}
  {{Early Results from GLASS-JWST. III. Galaxy Candidates at z 9-15}},\ }\href {https://doi.org/10.3847/2041-8213/ac94d0} {\bibfield  {journal} {\bibinfo  {journal} {\apjl}\ }\textbf {\bibinfo {volume} {938}},\ \bibinfo {eid} {L15} (\bibinfo {year} {2022})},\ \Eprint {https://arxiv.org/abs/2207.09436} {arXiv:2207.09436 [astro-ph.GA]} \BibitemShut {NoStop}%
\bibitem [{\citenamefont {{Finkelstein}}\ \emph {et~al.}(2022)\citenamefont {{Finkelstein}}, \citenamefont {{Bagley}}, \citenamefont {{Haro}}, \citenamefont {{Dickinson}}, \citenamefont {{Ferguson}}, \citenamefont {{Kartaltepe}}, \citenamefont {{Papovich}}, \citenamefont {{Burgarella}}, \citenamefont {{Kocevski}}, \citenamefont {{Huertas-Company}}, \citenamefont {{Iyer}}, \citenamefont {{Koekemoer}}, \citenamefont {{Larson}}, \citenamefont {{P{\'e}rez-Gonz{\'a}lez}}, \citenamefont {{Rose}}, \citenamefont {{Tacchella}}, \citenamefont {{Wilkins}}, \citenamefont {{Chworowsky}}, \citenamefont {{Medrano}}, \citenamefont {{Morales}}, \citenamefont {{Somerville}}, \citenamefont {{Yung}}, \citenamefont {{Fontana}}, \citenamefont {{Giavalisco}}, \citenamefont {{Grazian}}, \citenamefont {{Grogin}}, \citenamefont {{Kewley}}, \citenamefont {{Kirkpatrick}}, \citenamefont {{Kurczynski}}, \citenamefont {{Lotz}}, \citenamefont {{Pentericci}}, \citenamefont {{Pirzkal}}, \citenamefont {{Ravindranath}}, \citenamefont {{Ryan}},
  \citenamefont {{Trump}}, \citenamefont {{Yang}}, \citenamefont {{Almaini}}, \citenamefont {{Amor{\'\i}n}}, \citenamefont {{Annunziatella}}, \citenamefont {{Backhaus}}, \citenamefont {{Barro}}, \citenamefont {{Behroozi}}, \citenamefont {{Bell}}, \citenamefont {{Bhatawdekar}}, \citenamefont {{Bisigello}}, \citenamefont {{Bromm}}, \citenamefont {{Buat}}, \citenamefont {{Buitrago}}, \citenamefont {{Calabr{\`o}}}, \citenamefont {{Casey}}, \citenamefont {{Castellano}}, \citenamefont {{Ch{\'a}vez Ortiz}}, \citenamefont {{Ciesla}}, \citenamefont {{Cleri}}, \citenamefont {{Cohen}}, \citenamefont {{Cole}}, \citenamefont {{Cooke}}, \citenamefont {{Cooper}}, \citenamefont {{Cooray}}, \citenamefont {{Costantin}}, \citenamefont {{Cox}}, \citenamefont {{Croton}}, \citenamefont {{Daddi}}, \citenamefont {{Dav{\'e}}}, \citenamefont {{de La Vega}}, \citenamefont {{Dekel}}, \citenamefont {{Elbaz}}, \citenamefont {{Estrada-Carpenter}}, \citenamefont {{Faber}}, \citenamefont {{Fern{\'a}ndez}}, \citenamefont {{Finkelstein}},
  \citenamefont {{Freundlich}}, \citenamefont {{Fujimoto}}, \citenamefont {{Garc{\'\i}a-Argum{\'a}nez}}, \citenamefont {{Gardner}}, \citenamefont {{Gawiser}}, \citenamefont {{G{\'o}mez-Guijarro}}, \citenamefont {{Guo}}, \citenamefont {{Hamblin}}, \citenamefont {{Hamilton}}, \citenamefont {{Hathi}}, \citenamefont {{Holwerda}}, \citenamefont {{Hirschmann}}, \citenamefont {{Hutchison}}, \citenamefont {{Jaskot}}, \citenamefont {{Jha}}, \citenamefont {{Jogee}}, \citenamefont {{Juneau}}, \citenamefont {{Jung}}, \citenamefont {{Kassin}}, \citenamefont {{Bail}}, \citenamefont {{Leung}}, \citenamefont {{Lucas}}, \citenamefont {{Magnelli}}, \citenamefont {{Mantha}}, \citenamefont {{Matharu}}, \citenamefont {{McGrath}}, \citenamefont {{McIntosh}}, \citenamefont {{Merlin}}, \citenamefont {{Mobasher}}, \citenamefont {{Newman}}, \citenamefont {{Nicholls}}, \citenamefont {{Pandya}}, \citenamefont {{Rafelski}}, \citenamefont {{Ronayne}}, \citenamefont {{Santini}}, \citenamefont {{Seill{\'e}}}, \citenamefont {{Shah}},
  \citenamefont {{Shen}}, \citenamefont {{Simons}}, \citenamefont {{Snyder}}, \citenamefont {{Stanway}}, \citenamefont {{Straughn}}, \citenamefont {{Teplitz}}, \citenamefont {{Vanderhoof}}, \citenamefont {{Vega-Ferrero}}, \citenamefont {{Wang}}, \citenamefont {{Weiner}}, \citenamefont {{Willmer}}, \citenamefont {{Wuyts}}, \citenamefont {{Zavala}},\ and\ \citenamefont {{Ceers Team}}}]{Finkelstein22}%
  \BibitemOpen
  \bibfield  {author} {\bibinfo {author} {\bibfnamefont {S.~L.}\ \bibnamefont {{Finkelstein}}}, \bibinfo {author} {\bibfnamefont {M.~B.}\ \bibnamefont {{Bagley}}}, \bibinfo {author} {\bibfnamefont {P.~A.}\ \bibnamefont {{Haro}}}, \bibinfo {author} {\bibfnamefont {M.}~\bibnamefont {{Dickinson}}}, \bibinfo {author} {\bibfnamefont {H.~C.}\ \bibnamefont {{Ferguson}}}, \bibinfo {author} {\bibfnamefont {J.~S.}\ \bibnamefont {{Kartaltepe}}}, \bibinfo {author} {\bibfnamefont {C.}~\bibnamefont {{Papovich}}}, \bibinfo {author} {\bibfnamefont {D.}~\bibnamefont {{Burgarella}}}, \bibinfo {author} {\bibfnamefont {D.~D.}\ \bibnamefont {{Kocevski}}}, \bibinfo {author} {\bibfnamefont {M.}~\bibnamefont {{Huertas-Company}}}, \bibinfo {author} {\bibfnamefont {K.~G.}\ \bibnamefont {{Iyer}}}, \bibinfo {author} {\bibfnamefont {A.~M.}\ \bibnamefont {{Koekemoer}}}, \bibinfo {author} {\bibfnamefont {R.~L.}\ \bibnamefont {{Larson}}}, \bibinfo {author} {\bibfnamefont {P.~G.}\ \bibnamefont {{P{\'e}rez-Gonz{\'a}lez}}}, \bibinfo {author}
  {\bibfnamefont {C.}~\bibnamefont {{Rose}}}, \bibinfo {author} {\bibfnamefont {S.}~\bibnamefont {{Tacchella}}}, \bibinfo {author} {\bibfnamefont {S.~M.}\ \bibnamefont {{Wilkins}}}, \bibinfo {author} {\bibfnamefont {K.}~\bibnamefont {{Chworowsky}}}, \bibinfo {author} {\bibfnamefont {A.}~\bibnamefont {{Medrano}}}, \bibinfo {author} {\bibfnamefont {A.~M.}\ \bibnamefont {{Morales}}}, \bibinfo {author} {\bibfnamefont {R.~S.}\ \bibnamefont {{Somerville}}}, \bibinfo {author} {\bibfnamefont {L.~Y.~A.}\ \bibnamefont {{Yung}}}, \bibinfo {author} {\bibfnamefont {A.}~\bibnamefont {{Fontana}}}, \bibinfo {author} {\bibfnamefont {M.}~\bibnamefont {{Giavalisco}}}, \bibinfo {author} {\bibfnamefont {A.}~\bibnamefont {{Grazian}}}, \bibinfo {author} {\bibfnamefont {N.~A.}\ \bibnamefont {{Grogin}}}, \bibinfo {author} {\bibfnamefont {L.~J.}\ \bibnamefont {{Kewley}}}, \bibinfo {author} {\bibfnamefont {A.}~\bibnamefont {{Kirkpatrick}}}, \bibinfo {author} {\bibfnamefont {P.}~\bibnamefont {{Kurczynski}}}, \bibinfo {author}
  {\bibfnamefont {J.~M.}\ \bibnamefont {{Lotz}}}, \bibinfo {author} {\bibfnamefont {L.}~\bibnamefont {{Pentericci}}}, \bibinfo {author} {\bibfnamefont {N.}~\bibnamefont {{Pirzkal}}}, \bibinfo {author} {\bibfnamefont {S.}~\bibnamefont {{Ravindranath}}}, \bibinfo {author} {\bibfnamefont {R.~E.}\ \bibnamefont {{Ryan}}}, \bibinfo {author} {\bibfnamefont {J.~R.}\ \bibnamefont {{Trump}}}, \bibinfo {author} {\bibfnamefont {G.}~\bibnamefont {{Yang}}}, \bibinfo {author} {\bibfnamefont {O.}~\bibnamefont {{Almaini}}}, \bibinfo {author} {\bibfnamefont {R.~O.}\ \bibnamefont {{Amor{\'\i}n}}}, \bibinfo {author} {\bibfnamefont {M.}~\bibnamefont {{Annunziatella}}}, \bibinfo {author} {\bibfnamefont {B.~E.}\ \bibnamefont {{Backhaus}}}, \bibinfo {author} {\bibfnamefont {G.}~\bibnamefont {{Barro}}}, \bibinfo {author} {\bibfnamefont {P.}~\bibnamefont {{Behroozi}}}, \bibinfo {author} {\bibfnamefont {E.~F.}\ \bibnamefont {{Bell}}}, \bibinfo {author} {\bibfnamefont {R.}~\bibnamefont {{Bhatawdekar}}}, \bibinfo {author} {\bibfnamefont
  {L.}~\bibnamefont {{Bisigello}}}, \bibinfo {author} {\bibfnamefont {V.}~\bibnamefont {{Bromm}}}, \bibinfo {author} {\bibfnamefont {V.}~\bibnamefont {{Buat}}}, \bibinfo {author} {\bibfnamefont {F.}~\bibnamefont {{Buitrago}}}, \bibinfo {author} {\bibfnamefont {A.}~\bibnamefont {{Calabr{\`o}}}}, \bibinfo {author} {\bibfnamefont {C.~M.}\ \bibnamefont {{Casey}}}, \bibinfo {author} {\bibfnamefont {M.}~\bibnamefont {{Castellano}}}, \bibinfo {author} {\bibfnamefont {{\'O}.~A.}\ \bibnamefont {{Ch{\'a}vez Ortiz}}}, \bibinfo {author} {\bibfnamefont {L.}~\bibnamefont {{Ciesla}}}, \bibinfo {author} {\bibfnamefont {N.~J.}\ \bibnamefont {{Cleri}}}, \bibinfo {author} {\bibfnamefont {S.~H.}\ \bibnamefont {{Cohen}}}, \bibinfo {author} {\bibfnamefont {J.~W.}\ \bibnamefont {{Cole}}}, \bibinfo {author} {\bibfnamefont {K.~C.}\ \bibnamefont {{Cooke}}}, \bibinfo {author} {\bibfnamefont {M.~C.}\ \bibnamefont {{Cooper}}}, \bibinfo {author} {\bibfnamefont {A.~R.}\ \bibnamefont {{Cooray}}}, \bibinfo {author} {\bibfnamefont
  {L.}~\bibnamefont {{Costantin}}}, \bibinfo {author} {\bibfnamefont {I.~G.}\ \bibnamefont {{Cox}}}, \bibinfo {author} {\bibfnamefont {D.}~\bibnamefont {{Croton}}}, \bibinfo {author} {\bibfnamefont {E.}~\bibnamefont {{Daddi}}}, \bibinfo {author} {\bibfnamefont {R.}~\bibnamefont {{Dav{\'e}}}}, \bibinfo {author} {\bibfnamefont {A.}~\bibnamefont {{de La Vega}}}, \bibinfo {author} {\bibfnamefont {A.}~\bibnamefont {{Dekel}}}, \bibinfo {author} {\bibfnamefont {D.}~\bibnamefont {{Elbaz}}}, \bibinfo {author} {\bibfnamefont {V.}~\bibnamefont {{Estrada-Carpenter}}}, \bibinfo {author} {\bibfnamefont {S.~M.}\ \bibnamefont {{Faber}}}, \bibinfo {author} {\bibfnamefont {V.}~\bibnamefont {{Fern{\'a}ndez}}}, \bibinfo {author} {\bibfnamefont {K.~D.}\ \bibnamefont {{Finkelstein}}}, \bibinfo {author} {\bibfnamefont {J.}~\bibnamefont {{Freundlich}}}, \bibinfo {author} {\bibfnamefont {S.}~\bibnamefont {{Fujimoto}}}, \bibinfo {author} {\bibfnamefont {{\'A}.}~\bibnamefont {{Garc{\'\i}a-Argum{\'a}nez}}}, \bibinfo {author}
  {\bibfnamefont {J.~P.}\ \bibnamefont {{Gardner}}}, \bibinfo {author} {\bibfnamefont {E.}~\bibnamefont {{Gawiser}}}, \bibinfo {author} {\bibfnamefont {C.}~\bibnamefont {{G{\'o}mez-Guijarro}}}, \bibinfo {author} {\bibfnamefont {Y.}~\bibnamefont {{Guo}}}, \bibinfo {author} {\bibfnamefont {K.}~\bibnamefont {{Hamblin}}}, \bibinfo {author} {\bibfnamefont {T.~S.}\ \bibnamefont {{Hamilton}}}, \bibinfo {author} {\bibfnamefont {N.~P.}\ \bibnamefont {{Hathi}}}, \bibinfo {author} {\bibfnamefont {B.~W.}\ \bibnamefont {{Holwerda}}}, \bibinfo {author} {\bibfnamefont {M.}~\bibnamefont {{Hirschmann}}}, \bibinfo {author} {\bibfnamefont {T.~A.}\ \bibnamefont {{Hutchison}}}, \bibinfo {author} {\bibfnamefont {A.~E.}\ \bibnamefont {{Jaskot}}}, \bibinfo {author} {\bibfnamefont {S.~W.}\ \bibnamefont {{Jha}}}, \bibinfo {author} {\bibfnamefont {S.}~\bibnamefont {{Jogee}}}, \bibinfo {author} {\bibfnamefont {S.}~\bibnamefont {{Juneau}}}, \bibinfo {author} {\bibfnamefont {I.}~\bibnamefont {{Jung}}}, \bibinfo {author} {\bibfnamefont
  {S.~A.}\ \bibnamefont {{Kassin}}}, \bibinfo {author} {\bibfnamefont {A.~L.}\ \bibnamefont {{Bail}}}, \bibinfo {author} {\bibfnamefont {G.~C.~K.}\ \bibnamefont {{Leung}}}, \bibinfo {author} {\bibfnamefont {R.~A.}\ \bibnamefont {{Lucas}}}, \bibinfo {author} {\bibfnamefont {B.}~\bibnamefont {{Magnelli}}}, \bibinfo {author} {\bibfnamefont {K.~B.}\ \bibnamefont {{Mantha}}}, \bibinfo {author} {\bibfnamefont {J.}~\bibnamefont {{Matharu}}}, \bibinfo {author} {\bibfnamefont {E.~J.}\ \bibnamefont {{McGrath}}}, \bibinfo {author} {\bibfnamefont {D.~H.}\ \bibnamefont {{McIntosh}}}, \bibinfo {author} {\bibfnamefont {E.}~\bibnamefont {{Merlin}}}, \bibinfo {author} {\bibfnamefont {B.}~\bibnamefont {{Mobasher}}}, \bibinfo {author} {\bibfnamefont {J.~A.}\ \bibnamefont {{Newman}}}, \bibinfo {author} {\bibfnamefont {D.~C.}\ \bibnamefont {{Nicholls}}}, \bibinfo {author} {\bibfnamefont {V.}~\bibnamefont {{Pandya}}}, \bibinfo {author} {\bibfnamefont {M.}~\bibnamefont {{Rafelski}}}, \bibinfo {author} {\bibfnamefont
  {K.}~\bibnamefont {{Ronayne}}}, \bibinfo {author} {\bibfnamefont {P.}~\bibnamefont {{Santini}}}, \bibinfo {author} {\bibfnamefont {L.-M.}\ \bibnamefont {{Seill{\'e}}}}, \bibinfo {author} {\bibfnamefont {E.~A.}\ \bibnamefont {{Shah}}}, \bibinfo {author} {\bibfnamefont {L.}~\bibnamefont {{Shen}}}, \bibinfo {author} {\bibfnamefont {R.~C.}\ \bibnamefont {{Simons}}}, \bibinfo {author} {\bibfnamefont {G.~F.}\ \bibnamefont {{Snyder}}}, \bibinfo {author} {\bibfnamefont {E.~R.}\ \bibnamefont {{Stanway}}}, \bibinfo {author} {\bibfnamefont {A.~N.}\ \bibnamefont {{Straughn}}}, \bibinfo {author} {\bibfnamefont {H.~I.}\ \bibnamefont {{Teplitz}}}, \bibinfo {author} {\bibfnamefont {B.~N.}\ \bibnamefont {{Vanderhoof}}}, \bibinfo {author} {\bibfnamefont {J.}~\bibnamefont {{Vega-Ferrero}}}, \bibinfo {author} {\bibfnamefont {W.}~\bibnamefont {{Wang}}}, \bibinfo {author} {\bibfnamefont {B.~J.}\ \bibnamefont {{Weiner}}}, \bibinfo {author} {\bibfnamefont {C.~N.~A.}\ \bibnamefont {{Willmer}}}, \bibinfo {author} {\bibfnamefont
  {S.}~\bibnamefont {{Wuyts}}}, \bibinfo {author} {\bibfnamefont {J.~A.}\ \bibnamefont {{Zavala}}},\ and\ \bibinfo {author} {\bibnamefont {{Ceers Team}}},\ }\bibfield  {title} {\bibinfo {title} {{A Long Time Ago in a Galaxy Far, Far Away: A Candidate z {\ensuremath{\sim}} 12 Galaxy in Early JWST CEERS Imaging}},\ }\href {https://doi.org/10.3847/2041-8213/ac966e} {\bibfield  {journal} {\bibinfo  {journal} {\apjl}\ }\textbf {\bibinfo {volume} {940}},\ \bibinfo {eid} {L55} (\bibinfo {year} {2022})},\ \Eprint {https://arxiv.org/abs/2207.12474} {arXiv:2207.12474 [astro-ph.GA]} \BibitemShut {NoStop}%
\bibitem [{\citenamefont {{Donnan}}\ \emph {et~al.}(2023)\citenamefont {{Donnan}}, \citenamefont {{McLeod}}, \citenamefont {{Dunlop}}, \citenamefont {{McLure}}, \citenamefont {{Carnall}}, \citenamefont {{Begley}}, \citenamefont {{Cullen}}, \citenamefont {{Hamadouche}}, \citenamefont {{Bowler}}, \citenamefont {{Magee}}, \citenamefont {{McCracken}}, \citenamefont {{Milvang-Jensen}}, \citenamefont {{Moneti}},\ and\ \citenamefont {{Targett}}}]{Donnan23}%
  \BibitemOpen
  \bibfield  {author} {\bibinfo {author} {\bibfnamefont {C.~T.}\ \bibnamefont {{Donnan}}}, \bibinfo {author} {\bibfnamefont {D.~J.}\ \bibnamefont {{McLeod}}}, \bibinfo {author} {\bibfnamefont {J.~S.}\ \bibnamefont {{Dunlop}}}, \bibinfo {author} {\bibfnamefont {R.~J.}\ \bibnamefont {{McLure}}}, \bibinfo {author} {\bibfnamefont {A.~C.}\ \bibnamefont {{Carnall}}}, \bibinfo {author} {\bibfnamefont {R.}~\bibnamefont {{Begley}}}, \bibinfo {author} {\bibfnamefont {F.}~\bibnamefont {{Cullen}}}, \bibinfo {author} {\bibfnamefont {M.~L.}\ \bibnamefont {{Hamadouche}}}, \bibinfo {author} {\bibfnamefont {R.~A.~A.}\ \bibnamefont {{Bowler}}}, \bibinfo {author} {\bibfnamefont {D.}~\bibnamefont {{Magee}}}, \bibinfo {author} {\bibfnamefont {H.~J.}\ \bibnamefont {{McCracken}}}, \bibinfo {author} {\bibfnamefont {B.}~\bibnamefont {{Milvang-Jensen}}}, \bibinfo {author} {\bibfnamefont {A.}~\bibnamefont {{Moneti}}},\ and\ \bibinfo {author} {\bibfnamefont {T.}~\bibnamefont {{Targett}}},\ }\href {https://doi.org/10.1093/mnras/stac3472}
  {\bibfield  {journal} {\bibinfo  {journal} {\mnras}\ }\textbf {\bibinfo {volume} {518}},\ \bibinfo {pages} {6011} (\bibinfo {year} {2023})},\ \Eprint {https://arxiv.org/abs/2207.12356} {arXiv:2207.12356 [astro-ph.GA]} \BibitemShut {NoStop}%
\bibitem [{\citenamefont {{Labb{\'e}}}\ \emph {et~al.}(2023)\citenamefont {{Labb{\'e}}}, \citenamefont {{van Dokkum}}, \citenamefont {{Nelson}}, \citenamefont {{Bezanson}}, \citenamefont {{Suess}}, \citenamefont {{Leja}}, \citenamefont {{Brammer}}, \citenamefont {{Whitaker}}, \citenamefont {{Mathews}}, \citenamefont {{Stefanon}},\ and\ \citenamefont {{Wang}}}]{Labbe23}%
  \BibitemOpen
  \bibfield  {author} {\bibinfo {author} {\bibfnamefont {I.}~\bibnamefont {{Labb{\'e}}}}, \bibinfo {author} {\bibfnamefont {P.}~\bibnamefont {{van Dokkum}}}, \bibinfo {author} {\bibfnamefont {E.}~\bibnamefont {{Nelson}}}, \bibinfo {author} {\bibfnamefont {R.}~\bibnamefont {{Bezanson}}}, \bibinfo {author} {\bibfnamefont {K.~A.}\ \bibnamefont {{Suess}}}, \bibinfo {author} {\bibfnamefont {J.}~\bibnamefont {{Leja}}}, \bibinfo {author} {\bibfnamefont {G.}~\bibnamefont {{Brammer}}}, \bibinfo {author} {\bibfnamefont {K.}~\bibnamefont {{Whitaker}}}, \bibinfo {author} {\bibfnamefont {E.}~\bibnamefont {{Mathews}}}, \bibinfo {author} {\bibfnamefont {M.}~\bibnamefont {{Stefanon}}},\ and\ \bibinfo {author} {\bibfnamefont {B.}~\bibnamefont {{Wang}}},\ }\bibfield  {title} {\bibinfo {title} {{A population of red candidate massive galaxies 600 Myr after the Big Bang}},\ }\href {https://doi.org/10.1038/s41586-023-05786-2} {\bibfield  {journal} {\bibinfo  {journal} {\nat}\ }\textbf {\bibinfo {volume} {616}},\ \bibinfo {pages}
  {266} (\bibinfo {year} {2023})},\ \Eprint {https://arxiv.org/abs/2207.12446} {arXiv:2207.12446 [astro-ph.GA]} \BibitemShut {NoStop}%
\bibitem [{\citenamefont {{Li}}\ \emph {et~al.}(2025)\citenamefont {{Li}}, \citenamefont {{Silverman}}, \citenamefont {{Shen}}, \citenamefont {{Volonteri}}, \citenamefont {{Jahnke}}, \citenamefont {{Zhuang}}, \citenamefont {{Scoggins}}, \citenamefont {{Ding}}, \citenamefont {{Harikane}}, \citenamefont {{Onoue}},\ and\ \citenamefont {{Tanaka}}}]{Li_2025_ApJ_981}%
  \BibitemOpen
  \bibfield  {author} {\bibinfo {author} {\bibfnamefont {J.}~\bibnamefont {{Li}}}, \bibinfo {author} {\bibfnamefont {J.~D.}\ \bibnamefont {{Silverman}}}, \bibinfo {author} {\bibfnamefont {Y.}~\bibnamefont {{Shen}}}, \bibinfo {author} {\bibfnamefont {M.}~\bibnamefont {{Volonteri}}}, \bibinfo {author} {\bibfnamefont {K.}~\bibnamefont {{Jahnke}}}, \bibinfo {author} {\bibfnamefont {M.-Y.}\ \bibnamefont {{Zhuang}}}, \bibinfo {author} {\bibfnamefont {M.~T.}\ \bibnamefont {{Scoggins}}}, \bibinfo {author} {\bibfnamefont {X.}~\bibnamefont {{Ding}}}, \bibinfo {author} {\bibfnamefont {Y.}~\bibnamefont {{Harikane}}}, \bibinfo {author} {\bibfnamefont {M.}~\bibnamefont {{Onoue}}},\ and\ \bibinfo {author} {\bibfnamefont {T.~S.}\ \bibnamefont {{Tanaka}}},\ }\bibfield  {title} {\bibinfo {title} {{Tip of the Iceberg: Overmassive Black Holes at $4 < z < 7$ Found by JWST Are Not Inconsistent with the Local $M_{BH}$ --- $M_\star$ Relation}},\ }\href {https://doi.org/10.3847/1538-4357/ada603} {\bibfield  {journal} {\bibinfo
  {journal} {\apj}\ }\textbf {\bibinfo {volume} {981}},\ \bibinfo {eid} {19} (\bibinfo {year} {2025})},\ \Eprint {https://arxiv.org/abs/2403.00074} {arXiv:2403.00074 [astro-ph.GA]} \BibitemShut {NoStop}%
\bibitem [{\citenamefont {{Akins}}\ \emph {et~al.}(2025)\citenamefont {{Akins}}, \citenamefont {{Casey}}, \citenamefont {{Berg}}, \citenamefont {{Chisholm}}, \citenamefont {{Cloonan}}, \citenamefont {{Franco}}, \citenamefont {{Finkelstein}}, \citenamefont {{Fujimoto}}, \citenamefont {{Koekemoer}}, \citenamefont {{Kokorev}}, \citenamefont {{Lambrides}}, \citenamefont {{Robertson}}, \citenamefont {{Taylor}}, \citenamefont {{Coulter}}, \citenamefont {{Fox}},\ and\ \citenamefont {{Karmen}}}]{Akins_2025}%
  \BibitemOpen
  \bibfield  {author} {\bibinfo {author} {\bibfnamefont {H.~B.}\ \bibnamefont {{Akins}}}, \bibinfo {author} {\bibfnamefont {C.~M.}\ \bibnamefont {{Casey}}}, \bibinfo {author} {\bibfnamefont {D.~A.}\ \bibnamefont {{Berg}}}, \bibinfo {author} {\bibfnamefont {J.}~\bibnamefont {{Chisholm}}}, \bibinfo {author} {\bibfnamefont {A.~P.}\ \bibnamefont {{Cloonan}}}, \bibinfo {author} {\bibfnamefont {M.}~\bibnamefont {{Franco}}}, \bibinfo {author} {\bibfnamefont {S.~L.}\ \bibnamefont {{Finkelstein}}}, \bibinfo {author} {\bibfnamefont {S.}~\bibnamefont {{Fujimoto}}}, \bibinfo {author} {\bibfnamefont {A.~M.}\ \bibnamefont {{Koekemoer}}}, \bibinfo {author} {\bibfnamefont {V.}~\bibnamefont {{Kokorev}}}, \bibinfo {author} {\bibfnamefont {E.}~\bibnamefont {{Lambrides}}}, \bibinfo {author} {\bibfnamefont {B.~E.}\ \bibnamefont {{Robertson}}}, \bibinfo {author} {\bibfnamefont {A.~J.}\ \bibnamefont {{Taylor}}}, \bibinfo {author} {\bibfnamefont {D.~A.}\ \bibnamefont {{Coulter}}}, \bibinfo {author} {\bibfnamefont {O.}~\bibnamefont
  {{Fox}}},\ and\ \bibinfo {author} {\bibfnamefont {M.}~\bibnamefont {{Karmen}}},\ }\bibfield  {title} {\bibinfo {title} {{Strong Rest-UV Emission Lines in a ``Little Red Dot'' Active Galactic Nucleus at z = 7: Early Supermassive Black Hole Growth alongside Compact Massive Star Formation?}},\ }\href {https://doi.org/10.3847/2041-8213/adab76} {\bibfield  {journal} {\bibinfo  {journal} {\apjl}\ }\textbf {\bibinfo {volume} {980}},\ \bibinfo {eid} {L29} (\bibinfo {year} {2025})},\ \Eprint {https://arxiv.org/abs/2410.00949} {arXiv:2410.00949 [astro-ph.GA]} \BibitemShut {NoStop}%
\bibitem [{\citenamefont {{Boylan-Kolchin}}(2023)}]{Boylan-Kolchin23}%
  \BibitemOpen
  \bibfield  {author} {\bibinfo {author} {\bibfnamefont {M.}~\bibnamefont {{Boylan-Kolchin}}},\ }\bibfield  {title} {\bibinfo {title} {{Stress testing {\ensuremath{\Lambda}}CDM with high-redshift galaxy candidates}},\ }\href {https://doi.org/10.1038/s41550-023-01937-7} {\bibfield  {journal} {\bibinfo  {journal} {Nature Astronomy}\ }\textbf {\bibinfo {volume} {7}},\ \bibinfo {pages} {731} (\bibinfo {year} {2023})},\ \Eprint {https://arxiv.org/abs/2208.01611} {arXiv:2208.01611 [astro-ph.CO]} \BibitemShut {NoStop}%
\bibitem [{\citenamefont {{Lovell}}\ \emph {et~al.}(2023)\citenamefont {{Lovell}}, \citenamefont {{Harrison}}, \citenamefont {{Harikane}}, \citenamefont {{Tacchella}},\ and\ \citenamefont {{Wilkins}}}]{Lovell22}%
  \BibitemOpen
  \bibfield  {author} {\bibinfo {author} {\bibfnamefont {C.~C.}\ \bibnamefont {{Lovell}}}, \bibinfo {author} {\bibfnamefont {I.}~\bibnamefont {{Harrison}}}, \bibinfo {author} {\bibfnamefont {Y.}~\bibnamefont {{Harikane}}}, \bibinfo {author} {\bibfnamefont {S.}~\bibnamefont {{Tacchella}}},\ and\ \bibinfo {author} {\bibfnamefont {S.~M.}\ \bibnamefont {{Wilkins}}},\ }\bibfield  {title} {\bibinfo {title} {{Extreme value statistics of the halo and stellar mass distributions at high redshift: are JWST results in tension with {\ensuremath{\Lambda}}CDM?}},\ }\href {https://doi.org/10.1093/mnras/stac3224} {\bibfield  {journal} {\bibinfo  {journal} {\mnras}\ }\textbf {\bibinfo {volume} {518}},\ \bibinfo {pages} {2511} (\bibinfo {year} {2023})},\ \Eprint {https://arxiv.org/abs/2208.10479} {arXiv:2208.10479 [astro-ph.GA]} \BibitemShut {NoStop}%
\bibitem [{\citenamefont {{Chen}}\ \emph {et~al.}(2023)\citenamefont {{Chen}}, \citenamefont {{Mo}},\ and\ \citenamefont {{Wang}}}]{Chen23}%
  \BibitemOpen
  \bibfield  {author} {\bibinfo {author} {\bibfnamefont {Y.}~\bibnamefont {{Chen}}}, \bibinfo {author} {\bibfnamefont {H.~J.}\ \bibnamefont {{Mo}}},\ and\ \bibinfo {author} {\bibfnamefont {K.}~\bibnamefont {{Wang}}},\ }\bibfield  {title} {\bibinfo {title} {{Massive dark matter haloes at high redshift: implications for observations in the JWST era}},\ }\href {https://doi.org/10.1093/mnras/stad2866} {\bibfield  {journal} {\bibinfo  {journal} {\mnras}\ }\textbf {\bibinfo {volume} {526}},\ \bibinfo {pages} {2542} (\bibinfo {year} {2023})},\ \Eprint {https://arxiv.org/abs/2304.13890} {arXiv:2304.13890 [astro-ph.GA]} \BibitemShut {NoStop}%
\bibitem [{\citenamefont {{Prada}}\ \emph {et~al.}(2023)\citenamefont {{Prada}}, \citenamefont {{Behroozi}}, \citenamefont {{Ishiyama}}, \citenamefont {{Klypin}},\ and\ \citenamefont {{P{\'e}rez}}}]{Prada23}%
  \BibitemOpen
  \bibfield  {author} {\bibinfo {author} {\bibfnamefont {F.}~\bibnamefont {{Prada}}}, \bibinfo {author} {\bibfnamefont {P.}~\bibnamefont {{Behroozi}}}, \bibinfo {author} {\bibfnamefont {T.}~\bibnamefont {{Ishiyama}}}, \bibinfo {author} {\bibfnamefont {A.}~\bibnamefont {{Klypin}}},\ and\ \bibinfo {author} {\bibfnamefont {E.}~\bibnamefont {{P{\'e}rez}}},\ }\bibfield  {title} {\bibinfo {title} {{Confirmation of the standard cosmological model from red massive galaxies $\sim600$ Myr after the Big Bang}},\ }\href {https://doi.org/10.48550/arXiv.2304.11911} {\bibfield  {journal} {\bibinfo  {journal} {arXiv e-prints}\ ,\ \bibinfo {eid} {arXiv:2304.11911}} (\bibinfo {year} {2023})},\ \Eprint {https://arxiv.org/abs/2304.11911} {arXiv:2304.11911 [astro-ph.GA]} \BibitemShut {NoStop}%
\bibitem [{\citenamefont {{Shen}}\ \emph {et~al.}(2023)\citenamefont {{Shen}}, \citenamefont {{Vogelsberger}}, \citenamefont {{Boylan-Kolchin}}, \citenamefont {{Tacchella}},\ and\ \citenamefont {{Kannan}}}]{Shen23}%
  \BibitemOpen
  \bibfield  {author} {\bibinfo {author} {\bibfnamefont {X.}~\bibnamefont {{Shen}}}, \bibinfo {author} {\bibfnamefont {M.}~\bibnamefont {{Vogelsberger}}}, \bibinfo {author} {\bibfnamefont {M.}~\bibnamefont {{Boylan-Kolchin}}}, \bibinfo {author} {\bibfnamefont {S.}~\bibnamefont {{Tacchella}}},\ and\ \bibinfo {author} {\bibfnamefont {R.}~\bibnamefont {{Kannan}}},\ }\bibfield  {title} {\bibinfo {title} {{The impact of UV variability on the abundance of bright galaxies at z {\ensuremath{\geq}} 9}},\ }\href {https://doi.org/10.1093/mnras/stad2508} {\bibfield  {journal} {\bibinfo  {journal} {\mnras}\ }\textbf {\bibinfo {volume} {525}},\ \bibinfo {pages} {3254} (\bibinfo {year} {2023})},\ \Eprint {https://arxiv.org/abs/2305.05679} {arXiv:2305.05679 [astro-ph.GA]} \BibitemShut {NoStop}%
\bibitem [{\citenamefont {{H{\"u}tsi}}\ \emph {et~al.}(2023)\citenamefont {{H{\"u}tsi}}, \citenamefont {{Raidal}}, \citenamefont {{Urrutia}}, \citenamefont {{Vaskonen}},\ and\ \citenamefont {{Veerm{\"a}e}}}]{Hutsi23}%
  \BibitemOpen
  \bibfield  {author} {\bibinfo {author} {\bibfnamefont {G.}~\bibnamefont {{H{\"u}tsi}}}, \bibinfo {author} {\bibfnamefont {M.}~\bibnamefont {{Raidal}}}, \bibinfo {author} {\bibfnamefont {J.}~\bibnamefont {{Urrutia}}}, \bibinfo {author} {\bibfnamefont {V.}~\bibnamefont {{Vaskonen}}},\ and\ \bibinfo {author} {\bibfnamefont {H.}~\bibnamefont {{Veerm{\"a}e}}},\ }\bibfield  {title} {\bibinfo {title} {{Did JWST observe imprints of axion miniclusters or primordial black holes?}},\ }\href {https://doi.org/10.1103/PhysRevD.107.043502} {\bibfield  {journal} {\bibinfo  {journal} {\prd}\ }\textbf {\bibinfo {volume} {107}},\ \bibinfo {eid} {043502} (\bibinfo {year} {2023})},\ \Eprint {https://arxiv.org/abs/2211.02651} {arXiv:2211.02651 [astro-ph.CO]} \BibitemShut {NoStop}%
\bibitem [{\citenamefont {{Inomata}}\ \emph {et~al.}(2023)\citenamefont {{Inomata}}, \citenamefont {{Braglia}},\ and\ \citenamefont {{Chen}}}]{inomata}%
  \BibitemOpen
  \bibfield  {author} {\bibinfo {author} {\bibfnamefont {K.}~\bibnamefont {{Inomata}}}, \bibinfo {author} {\bibfnamefont {M.}~\bibnamefont {{Braglia}}},\ and\ \bibinfo {author} {\bibfnamefont {X.}~\bibnamefont {{Chen}}},\ }\bibfield  {title} {\bibinfo {title} {{Questions on calculation of primordial power spectrum with large spikes: the resonance model case}},\ }\href {https://doi.org/10.1088/1475-7516/2023/04/011} {\bibfield  {journal} {\bibinfo  {journal} {\jcap}\ }\textbf {\bibinfo {volume} {2023}},\ \bibinfo {eid} {011} (\bibinfo {year} {2023})},\ \Eprint {https://arxiv.org/abs/2211.02586} {arXiv:2211.02586 [astro-ph.CO]} \BibitemShut {NoStop}%
\bibitem [{\citenamefont {{Ralegankar}}\ \emph {et~al.}(2024)\citenamefont {{Ralegankar}}, \citenamefont {{Pavi{\v{c}}evi{\'c}}},\ and\ \citenamefont {{Viel}}}]{2024arXiv240214079R}%
  \BibitemOpen
  \bibfield  {author} {\bibinfo {author} {\bibfnamefont {P.}~\bibnamefont {{Ralegankar}}}, \bibinfo {author} {\bibfnamefont {M.}~\bibnamefont {{Pavi{\v{c}}evi{\'c}}}},\ and\ \bibinfo {author} {\bibfnamefont {M.}~\bibnamefont {{Viel}}},\ }\bibfield  {title} {\bibinfo {title} {{Primordial magnetic fields: consistent initial conditions and impact on high-z structures}},\ }\href {https://doi.org/10.1088/1475-7516/2024/07/027} {\bibfield  {journal} {\bibinfo  {journal} {\jcap}\ }\textbf {\bibinfo {volume} {2024}},\ \bibinfo {eid} {027} (\bibinfo {year} {2024})},\ \Eprint {https://arxiv.org/abs/2402.14079} {arXiv:2402.14079 [astro-ph.CO]} \BibitemShut {NoStop}%
\bibitem [{\citenamefont {{Lukash}}\ and\ \citenamefont {{Mikheeva}}(2025)}]{lukash25}%
  \BibitemOpen
  \bibfield  {author} {\bibinfo {author} {\bibfnamefont {V.~N.}\ \bibnamefont {{Lukash}}}\ and\ \bibinfo {author} {\bibfnamefont {E.~V.}\ \bibnamefont {{Mikheeva}}},\ }\bibfield  {title} {\bibinfo {title} {Cascade relaxation of the gravitating vacuum as a generator of the evolving universe},\ }\href@noop {} {\bibfield  {journal} {\bibinfo  {journal} {JETP Lett.}\ }\textbf {\bibinfo {volume} {121}},\ \bibinfo {pages} {399} (\bibinfo {year} {2025})}\BibitemShut {NoStop}%
\bibitem [{\citenamefont {{Hirano}}\ \emph {et~al.}(2015)\citenamefont {{Hirano}}, \citenamefont {{Zhu}}, \citenamefont {{Yoshida}}, \citenamefont {{Spergel}},\ and\ \citenamefont {{Yorke}}}]{hirano15}%
  \BibitemOpen
  \bibfield  {author} {\bibinfo {author} {\bibfnamefont {S.}~\bibnamefont {{Hirano}}}, \bibinfo {author} {\bibfnamefont {N.}~\bibnamefont {{Zhu}}}, \bibinfo {author} {\bibfnamefont {N.}~\bibnamefont {{Yoshida}}}, \bibinfo {author} {\bibfnamefont {D.}~\bibnamefont {{Spergel}}},\ and\ \bibinfo {author} {\bibfnamefont {H.~W.}\ \bibnamefont {{Yorke}}},\ }\bibfield  {title} {\bibinfo {title} {{Early Structure Formation from Primordial Density Fluctuations with a Blue, Tilted Power Spectrum}},\ }\href {https://doi.org/10.1088/0004-637X/814/1/18} {\bibfield  {journal} {\bibinfo  {journal} {\apj}\ }\textbf {\bibinfo {volume} {814}},\ \bibinfo {eid} {18} (\bibinfo {year} {2015})},\ \Eprint {https://arxiv.org/abs/1504.05186} {arXiv:1504.05186 [astro-ph.CO]} \BibitemShut {NoStop}%
\bibitem [{\citenamefont {{Parashari}}\ and\ \citenamefont {{Laha}}(2023)}]{parashari23}%
  \BibitemOpen
  \bibfield  {author} {\bibinfo {author} {\bibfnamefont {P.}~\bibnamefont {{Parashari}}}\ and\ \bibinfo {author} {\bibfnamefont {R.}~\bibnamefont {{Laha}}},\ }\bibfield  {title} {\bibinfo {title} {{Primordial power spectrum in light of JWST observations of high redshift galaxies}},\ }\href {https://doi.org/10.1093/mnrasl/slad107} {\bibfield  {journal} {\bibinfo  {journal} {\mnras}\ }\textbf {\bibinfo {volume} {526}},\ \bibinfo {pages} {L63} (\bibinfo {year} {2023})},\ \Eprint {https://arxiv.org/abs/2305.00999} {arXiv:2305.00999 [astro-ph.CO]} \BibitemShut {NoStop}%
\bibitem [{\citenamefont {Padmanabhan}\ and\ \citenamefont {Loeb}(2023)}]{padma23}%
  \BibitemOpen
  \bibfield  {author} {\bibinfo {author} {\bibfnamefont {H.}~\bibnamefont {Padmanabhan}}\ and\ \bibinfo {author} {\bibfnamefont {A.}~\bibnamefont {Loeb}},\ }\bibfield  {title} {\bibinfo {title} {{Alleviating the Need for Exponential Evolution of JWST Galaxies in $10^{10}$ M$\odot$ Haloes at z $>$ 10 by a Modified $\Lambda$CDM Power Spectrum}},\ }\href {https://doi.org/10.3847/2041-8213/acea7a} {\bibfield  {journal} {\bibinfo  {journal} {\apjl}\ }\textbf {\bibinfo {volume} {953}},\ \bibinfo {pages} {L4} (\bibinfo {year} {2023})}\BibitemShut {NoStop}%
\bibitem [{\citenamefont {{Tkachev}}\ \emph {et~al.}(2024{\natexlab{a}})\citenamefont {{Tkachev}}, \citenamefont {{Pilipenko}}, \citenamefont {{Mikheeva}},\ and\ \citenamefont {{Lukash}}}]{tkachev23}%
  \BibitemOpen
  \bibfield  {author} {\bibinfo {author} {\bibfnamefont {M.~V.}\ \bibnamefont {{Tkachev}}}, \bibinfo {author} {\bibfnamefont {S.~V.}\ \bibnamefont {{Pilipenko}}}, \bibinfo {author} {\bibfnamefont {E.~V.}\ \bibnamefont {{Mikheeva}}},\ and\ \bibinfo {author} {\bibfnamefont {V.~N.}\ \bibnamefont {{Lukash}}},\ }\bibfield  {title} {\bibinfo {title} {{Excess of high-$z$ galaxies as a test for bumpy power spectrum of density perturbations}},\ }\href {https://doi.org/10.1093/mnras/stad3279} {\bibfield  {journal} {\bibinfo  {journal} {\mnras}\ }\textbf {\bibinfo {volume} {527}},\ \bibinfo {pages} {1381} (\bibinfo {year} {2024}{\natexlab{a}})},\ \Eprint {https://arxiv.org/abs/2307.13774} {arXiv:2307.13774 [astro-ph.CO]} \BibitemShut {NoStop}%
\bibitem [{\citenamefont {{Hirano}}\ and\ \citenamefont {{Yoshida}}(2024)}]{hirano24}%
  \BibitemOpen
  \bibfield  {author} {\bibinfo {author} {\bibfnamefont {S.}~\bibnamefont {{Hirano}}}\ and\ \bibinfo {author} {\bibfnamefont {N.}~\bibnamefont {{Yoshida}}},\ }\bibfield  {title} {\bibinfo {title} {{Early Structure Formation from Primordial Density Fluctuations with a Blue, Tilted Power Spectrum: High-redshift Galaxies}},\ }\href {https://doi.org/10.3847/1538-4357/ad22e0} {\bibfield  {journal} {\bibinfo  {journal} {\apj}\ }\textbf {\bibinfo {volume} {963}},\ \bibinfo {eid} {2} (\bibinfo {year} {2024})},\ \Eprint {https://arxiv.org/abs/2306.11993} {arXiv:2306.11993 [astro-ph.GA]} \BibitemShut {NoStop}%
\bibitem [{\citenamefont {{Eroshenko}}\ \emph {et~al.}(2024)\citenamefont {{Eroshenko}}, \citenamefont {{Lukash}}, \citenamefont {{Mikheeva}}, \citenamefont {{Pilipenko}},\ and\ \citenamefont {{Tkachev}}}]{yura2024}%
  \BibitemOpen
  \bibfield  {author} {\bibinfo {author} {\bibfnamefont {Y.~N.}\ \bibnamefont {{Eroshenko}}}, \bibinfo {author} {\bibfnamefont {V.~N.}\ \bibnamefont {{Lukash}}}, \bibinfo {author} {\bibfnamefont {E.~V.}\ \bibnamefont {{Mikheeva}}}, \bibinfo {author} {\bibfnamefont {S.~V.}\ \bibnamefont {{Pilipenko}}},\ and\ \bibinfo {author} {\bibfnamefont {M.~V.}\ \bibnamefont {{Tkachev}}},\ }\bibfield  {title} {\bibinfo {title} {{Properties of Central Regions of the Dark Matter Halos in the Model with a Bump in the Power Spectrum of Density Perturbations}},\ }\href {https://doi.org/10.1134/S0021364024601775} {\bibfield  {journal} {\bibinfo  {journal} {JETP Lett.}\ }\textbf {\bibinfo {volume} {120}},\ \bibinfo {pages} {83} (\bibinfo {year} {2024})},\ \Eprint {https://arxiv.org/abs/2409.02739} {arXiv:2409.02739 [astro-ph.GA]} \BibitemShut {NoStop}%
\bibitem [{\citenamefont {{Starobinsky}}(1992)}]{star}%
  \BibitemOpen
  \bibfield  {author} {\bibinfo {author} {\bibfnamefont {A.~A.}\ \bibnamefont {{Starobinsky}}},\ }\bibfield  {title} {\bibinfo {title} {{Spectrum of adiabatic perturbations in the universe when there are singularities in the inflaton potential}},\ }\href@noop {} {\bibfield  {journal} {\bibinfo  {journal} {JETP Lett.}\ }\textbf {\bibinfo {volume} {55}},\ \bibinfo {eid} {489-494} (\bibinfo {year} {1992})}\BibitemShut {NoStop}%
\bibitem [{\citenamefont {{Ivanov}}\ \emph {et~al.}(1994)\citenamefont {{Ivanov}}, \citenamefont {{Naselsky}},\ and\ \citenamefont {{Novikov}}}]{inn}%
  \BibitemOpen
  \bibfield  {author} {\bibinfo {author} {\bibfnamefont {P.}~\bibnamefont {{Ivanov}}}, \bibinfo {author} {\bibfnamefont {P.}~\bibnamefont {{Naselsky}}},\ and\ \bibinfo {author} {\bibfnamefont {I.}~\bibnamefont {{Novikov}}},\ }\bibfield  {title} {\bibinfo {title} {{Inflation and primordial black holes as dark matter}},\ }\href {https://doi.org/10.1103/PhysRevD.50.7173} {\bibfield  {journal} {\bibinfo  {journal} {\prd}\ }\textbf {\bibinfo {volume} {50}},\ \bibinfo {pages} {7173} (\bibinfo {year} {1994})}\BibitemShut {NoStop}%
\bibitem [{\citenamefont {{Peng}}\ \emph {et~al.}(2021)\citenamefont {{Peng}}, \citenamefont {{Fu}}, \citenamefont {{Liu}}, \citenamefont {{Guo}},\ and\ \citenamefont {{Cai}}}]{ZZPeng}%
  \BibitemOpen
  \bibfield  {author} {\bibinfo {author} {\bibfnamefont {Z.-Z.}\ \bibnamefont {{Peng}}}, \bibinfo {author} {\bibfnamefont {C.}~\bibnamefont {{Fu}}}, \bibinfo {author} {\bibfnamefont {J.}~\bibnamefont {{Liu}}}, \bibinfo {author} {\bibfnamefont {Z.-K.}\ \bibnamefont {{Guo}}},\ and\ \bibinfo {author} {\bibfnamefont {R.-G.}\ \bibnamefont {{Cai}}},\ }\bibfield  {title} {\bibinfo {title} {{Gravitational waves from resonant amplification of curvature perturbations during inflation}},\ }\href {https://doi.org/10.1088/1475-7516/2021/10/050} {\bibfield  {journal} {\bibinfo  {journal} {\jcap}\ }\textbf {\bibinfo {volume} {2021}},\ \bibinfo {eid} {050} (\bibinfo {year} {2021})},\ \Eprint {https://arxiv.org/abs/2106.11816} {arXiv:2106.11816 [astro-ph.CO]} \BibitemShut {NoStop}%
\bibitem [{\citenamefont {{Lukash}}\ \emph {et~al.}(2000)\citenamefont {{Lukash}}, \citenamefont {{Mikheeva}}, \citenamefont {{M{\"u}ller}},\ and\ \citenamefont {{Malinovsky}}}]{lm2000a}%
  \BibitemOpen
  \bibfield  {author} {\bibinfo {author} {\bibfnamefont {V.~N.}\ \bibnamefont {{Lukash}}}, \bibinfo {author} {\bibfnamefont {E.~V.}\ \bibnamefont {{Mikheeva}}}, \bibinfo {author} {\bibfnamefont {V.}~\bibnamefont {{M{\"u}ller}}},\ and\ \bibinfo {author} {\bibfnamefont {A.~M.}\ \bibnamefont {{Malinovsky}}},\ }\bibfield  {title} {\bibinfo {title} {{Generalized inflation with a gravitational wave background}},\ }\href {https://doi.org/10.1046/j.1365-8711.2000.03663.x} {\bibfield  {journal} {\bibinfo  {journal} {\mnras}\ }\textbf {\bibinfo {volume} {317}},\ \bibinfo {pages} {795} (\bibinfo {year} {2000})},\ \Eprint {https://arxiv.org/abs/astro-ph/0005395} {arXiv:astro-ph/0005395 [astro-ph]} \BibitemShut {NoStop}%
\bibitem [{\citenamefont {{Lukash}}\ and\ \citenamefont {{Mikheeva}}(2000)}]{lm2000b}%
  \BibitemOpen
  \bibfield  {author} {\bibinfo {author} {\bibfnamefont {V.~N.}\ \bibnamefont {{Lukash}}}\ and\ \bibinfo {author} {\bibfnamefont {E.~V.}\ \bibnamefont {{Mikheeva}}},\ }\bibfield  {title} {\bibinfo {title} {{{\ensuremath{\Lambda}}-Inflation and CMB Anisotropy}},\ }\href {https://doi.org/10.1142/S0217751X00001798} {\bibfield  {journal} {\bibinfo  {journal} {International Journal of Modern Physics A}\ }\textbf {\bibinfo {volume} {15}},\ \bibinfo {pages} {3783} (\bibinfo {year} {2000})},\ \Eprint {https://arxiv.org/abs/astro-ph/9910135} {arXiv:astro-ph/9910135 [astro-ph]} \BibitemShut {NoStop}%
\bibitem [{\citenamefont {{Clesse}}\ and\ \citenamefont {{Garc{\'\i}a-Bellido}}(2015{\natexlab{a}})}]{Clesse2015}%
  \BibitemOpen
  \bibfield  {author} {\bibinfo {author} {\bibfnamefont {S.}~\bibnamefont {{Clesse}}}\ and\ \bibinfo {author} {\bibfnamefont {J.}~\bibnamefont {{Garc{\'\i}a-Bellido}}},\ }\bibfield  {title} {\bibinfo {title} {{Massive primordial black holes from hybrid inflation as dark matter and the seeds of galaxies}},\ }\href {https://doi.org/10.1103/PhysRevD.92.023524} {\bibfield  {journal} {\bibinfo  {journal} {\prd}\ }\textbf {\bibinfo {volume} {92}},\ \bibinfo {eid} {023524} (\bibinfo {year} {2015}{\natexlab{a}})},\ \Eprint {https://arxiv.org/abs/1501.07565} {arXiv:1501.07565 [astro-ph.CO]} \BibitemShut {NoStop}%
\bibitem [{\citenamefont {{Clesse}}\ and\ \citenamefont {{Garc{\'\i}a-Bellido}}(2015{\natexlab{b}})}]{GarciaMoralez2017}%
  \BibitemOpen
  \bibfield  {author} {\bibinfo {author} {\bibfnamefont {S.}~\bibnamefont {{Clesse}}}\ and\ \bibinfo {author} {\bibfnamefont {J.}~\bibnamefont {{Garc{\'\i}a-Bellido}}},\ }\bibfield  {title} {\bibinfo {title} {{Massive primordial black holes from hybrid inflation as dark matter and the seeds of galaxies}},\ }\href {https://doi.org/10.1103/PhysRevD.92.023524} {\bibfield  {journal} {\bibinfo  {journal} {\prd}\ }\textbf {\bibinfo {volume} {92}},\ \bibinfo {eid} {023524} (\bibinfo {year} {2015}{\natexlab{b}})},\ \Eprint {https://arxiv.org/abs/1501.07565} {arXiv:1501.07565 [astro-ph.CO]} \BibitemShut {NoStop}%
\bibitem [{\citenamefont {{Garc{\'\i}a-Bellido}}\ and\ \citenamefont {{Ruiz Morales}}(2017)}]{inflec1}%
  \BibitemOpen
  \bibfield  {author} {\bibinfo {author} {\bibfnamefont {J.}~\bibnamefont {{Garc{\'\i}a-Bellido}}}\ and\ \bibinfo {author} {\bibfnamefont {E.}~\bibnamefont {{Ruiz Morales}}},\ }\bibfield  {title} {\bibinfo {title} {{Primordial black holes from single field models of inflation}},\ }\href {https://doi.org/10.1016/j.dark.2017.09.007} {\bibfield  {journal} {\bibinfo  {journal} {Physics of the Dark Universe}\ }\textbf {\bibinfo {volume} {18}},\ \bibinfo {pages} {47} (\bibinfo {year} {2017})},\ \Eprint {https://arxiv.org/abs/1702.03901} {arXiv:1702.03901 [astro-ph.CO]} \BibitemShut {NoStop}%
\bibitem [{\citenamefont {{Germani}}\ and\ \citenamefont {{Prokopec}}(2017)}]{inflec2}%
  \BibitemOpen
  \bibfield  {author} {\bibinfo {author} {\bibfnamefont {C.}~\bibnamefont {{Germani}}}\ and\ \bibinfo {author} {\bibfnamefont {T.}~\bibnamefont {{Prokopec}}},\ }\bibfield  {title} {\bibinfo {title} {{On primordial black holes from an inflection point}},\ }\href {https://doi.org/10.1016/j.dark.2017.09.001} {\bibfield  {journal} {\bibinfo  {journal} {Physics of the Dark Universe}\ }\textbf {\bibinfo {volume} {18}},\ \bibinfo {pages} {6} (\bibinfo {year} {2017})},\ \Eprint {https://arxiv.org/abs/1706.04226} {arXiv:1706.04226 [astro-ph.CO]} \BibitemShut {NoStop}%
\bibitem [{\citenamefont {Gao}\ and\ \citenamefont {Guo}(2018)}]{inflec3}%
  \BibitemOpen
  \bibfield  {author} {\bibinfo {author} {\bibfnamefont {T.-J.}\ \bibnamefont {Gao}}\ and\ \bibinfo {author} {\bibfnamefont {Z.-K.}\ \bibnamefont {Guo}},\ }\bibfield  {title} {\bibinfo {title} {Primordial black hole production in inflationary models of supergravity with a single chiral superfield},\ }\bibfield  {journal} {\bibinfo  {journal} {Physical Review D}\ }\textbf {\bibinfo {volume} {98}},\ \href {https://doi.org/10.1103/physrevd.98.063526} {10.1103/physrevd.98.063526} (\bibinfo {year} {2018})\BibitemShut {NoStop}%
\bibitem [{\citenamefont {Xu}\ \emph {et~al.}(2020)\citenamefont {Xu}, \citenamefont {Liu}, \citenamefont {Gao},\ and\ \citenamefont {Guo}}]{inflec4}%
  \BibitemOpen
  \bibfield  {author} {\bibinfo {author} {\bibfnamefont {W.-T.}\ \bibnamefont {Xu}}, \bibinfo {author} {\bibfnamefont {J.}~\bibnamefont {Liu}}, \bibinfo {author} {\bibfnamefont {T.-J.}\ \bibnamefont {Gao}},\ and\ \bibinfo {author} {\bibfnamefont {Z.-K.}\ \bibnamefont {Guo}},\ }\bibfield  {title} {\bibinfo {title} {Gravitational waves from double-inflection-point inflation},\ }\bibfield  {journal} {\bibinfo  {journal} {Physical Review D}\ }\textbf {\bibinfo {volume} {101}},\ \href {https://doi.org/10.1103/physrevd.101.023505} {10.1103/physrevd.101.023505} (\bibinfo {year} {2020})\BibitemShut {NoStop}%
\bibitem [{\citenamefont {{Tkachev}}\ \emph {et~al.}(2024{\natexlab{b}})\citenamefont {{Tkachev}}, \citenamefont {{Pilipenko}}, \citenamefont {{Mikheeva}},\ and\ \citenamefont {{Lukash}}}]{PRD2024}%
  \BibitemOpen
  \bibfield  {author} {\bibinfo {author} {\bibfnamefont {M.~V.}\ \bibnamefont {{Tkachev}}}, \bibinfo {author} {\bibfnamefont {S.~V.}\ \bibnamefont {{Pilipenko}}}, \bibinfo {author} {\bibfnamefont {E.~V.}\ \bibnamefont {{Mikheeva}}},\ and\ \bibinfo {author} {\bibfnamefont {V.~N.}\ \bibnamefont {{Lukash}}},\ }\bibfield  {title} {\bibinfo {title} {{Inner structure of dark matter halos at high $z$ in cosmological models with non-power-law primordial spectra}},\ }\href {https://doi.org/10.1103/PhysRevD.110.083530} {\bibfield  {journal} {\bibinfo  {journal} {\prd}\ }\textbf {\bibinfo {volume} {110}},\ \bibinfo {eid} {083530} (\bibinfo {year} {2024}{\natexlab{b}})},\ \Eprint {https://arxiv.org/abs/2407.02991} {arXiv:2407.02991 [astro-ph.CO]} \BibitemShut {NoStop}%
\bibitem [{\citenamefont {{Silk}}\ \emph {et~al.}(2024)\citenamefont {{Silk}}, \citenamefont {{Begelman}}, \citenamefont {{Norman}}, \citenamefont {{Nusser}},\ and\ \citenamefont {{Wyse}}}]{SilkBegelman2024}%
  \BibitemOpen
  \bibfield  {author} {\bibinfo {author} {\bibfnamefont {J.}~\bibnamefont {{Silk}}}, \bibinfo {author} {\bibfnamefont {M.~C.}\ \bibnamefont {{Begelman}}}, \bibinfo {author} {\bibfnamefont {C.}~\bibnamefont {{Norman}}}, \bibinfo {author} {\bibfnamefont {A.}~\bibnamefont {{Nusser}}},\ and\ \bibinfo {author} {\bibfnamefont {R.~F.~G.}\ \bibnamefont {{Wyse}}},\ }\bibfield  {title} {\bibinfo {title} {{Which Came First: Supermassive Black Holes or Galaxies? Insights from JWST}},\ }\href {https://doi.org/10.3847/2041-8213/ad1bf0} {\bibfield  {journal} {\bibinfo  {journal} {\apjl}\ }\textbf {\bibinfo {volume} {961}},\ \bibinfo {eid} {L39} (\bibinfo {year} {2024})},\ \Eprint {https://arxiv.org/abs/2401.02482} {arXiv:2401.02482 [astro-ph.GA]} \BibitemShut {NoStop}%
\bibitem [{\citenamefont {{Harikane}}\ \emph {et~al.}(2025)\citenamefont {{Harikane}}, \citenamefont {{Inoue}}, \citenamefont {{Ellis}}, \citenamefont {{Ouchi}}, \citenamefont {{Nakazato}}, \citenamefont {{Yoshida}}, \citenamefont {{Ono}}, \citenamefont {{Sun}}, \citenamefont {{Sato}}, \citenamefont {{Ferrami}}, \citenamefont {{Fujimoto}}, \citenamefont {{Kashikawa}}, \citenamefont {{McLeod}}, \citenamefont {{P{\'e}rez-Gonz{\'a}lez}}, \citenamefont {{Sawicki}}, \citenamefont {{Sugahara}}, \citenamefont {{Xu}}, \citenamefont {{Yamanaka}}, \citenamefont {{Carnall}}, \citenamefont {{Cullen}}, \citenamefont {{Dunlop}}, \citenamefont {{Egami}}, \citenamefont {{Grogin}}, \citenamefont {{Isobe}}, \citenamefont {{Koekemoer}}, \citenamefont {{Laporte}}, \citenamefont {{Lee}}, \citenamefont {{Magee}}, \citenamefont {{Matsuo}}, \citenamefont {{Matsuoka}}, \citenamefont {{Mawatari}}, \citenamefont {{Nakajima}}, \citenamefont {{Nakane}}, \citenamefont {{Tamura}}, \citenamefont {{Umeda}},\ and\ \citenamefont
  {{Yanagisawa}}}]{Harikane2025}%
  \BibitemOpen
  \bibfield  {author} {\bibinfo {author} {\bibfnamefont {Y.}~\bibnamefont {{Harikane}}}, \bibinfo {author} {\bibfnamefont {A.~K.}\ \bibnamefont {{Inoue}}}, \bibinfo {author} {\bibfnamefont {R.~S.}\ \bibnamefont {{Ellis}}}, \bibinfo {author} {\bibfnamefont {M.}~\bibnamefont {{Ouchi}}}, \bibinfo {author} {\bibfnamefont {Y.}~\bibnamefont {{Nakazato}}}, \bibinfo {author} {\bibfnamefont {N.}~\bibnamefont {{Yoshida}}}, \bibinfo {author} {\bibfnamefont {Y.}~\bibnamefont {{Ono}}}, \bibinfo {author} {\bibfnamefont {F.}~\bibnamefont {{Sun}}}, \bibinfo {author} {\bibfnamefont {R.~A.}\ \bibnamefont {{Sato}}}, \bibinfo {author} {\bibfnamefont {G.}~\bibnamefont {{Ferrami}}}, \bibinfo {author} {\bibfnamefont {S.}~\bibnamefont {{Fujimoto}}}, \bibinfo {author} {\bibfnamefont {N.}~\bibnamefont {{Kashikawa}}}, \bibinfo {author} {\bibfnamefont {D.~J.}\ \bibnamefont {{McLeod}}}, \bibinfo {author} {\bibfnamefont {P.~G.}\ \bibnamefont {{P{\'e}rez-Gonz{\'a}lez}}}, \bibinfo {author} {\bibfnamefont {M.}~\bibnamefont {{Sawicki}}},
  \bibinfo {author} {\bibfnamefont {Y.}~\bibnamefont {{Sugahara}}}, \bibinfo {author} {\bibfnamefont {Y.}~\bibnamefont {{Xu}}}, \bibinfo {author} {\bibfnamefont {S.}~\bibnamefont {{Yamanaka}}}, \bibinfo {author} {\bibfnamefont {A.~C.}\ \bibnamefont {{Carnall}}}, \bibinfo {author} {\bibfnamefont {F.}~\bibnamefont {{Cullen}}}, \bibinfo {author} {\bibfnamefont {J.~S.}\ \bibnamefont {{Dunlop}}}, \bibinfo {author} {\bibfnamefont {E.}~\bibnamefont {{Egami}}}, \bibinfo {author} {\bibfnamefont {N.}~\bibnamefont {{Grogin}}}, \bibinfo {author} {\bibfnamefont {Y.}~\bibnamefont {{Isobe}}}, \bibinfo {author} {\bibfnamefont {A.~M.}\ \bibnamefont {{Koekemoer}}}, \bibinfo {author} {\bibfnamefont {N.}~\bibnamefont {{Laporte}}}, \bibinfo {author} {\bibfnamefont {C.-H.}\ \bibnamefont {{Lee}}}, \bibinfo {author} {\bibfnamefont {D.}~\bibnamefont {{Magee}}}, \bibinfo {author} {\bibfnamefont {H.}~\bibnamefont {{Matsuo}}}, \bibinfo {author} {\bibfnamefont {Y.}~\bibnamefont {{Matsuoka}}}, \bibinfo {author} {\bibfnamefont
  {K.}~\bibnamefont {{Mawatari}}}, \bibinfo {author} {\bibfnamefont {K.}~\bibnamefont {{Nakajima}}}, \bibinfo {author} {\bibfnamefont {M.}~\bibnamefont {{Nakane}}}, \bibinfo {author} {\bibfnamefont {Y.}~\bibnamefont {{Tamura}}}, \bibinfo {author} {\bibfnamefont {H.}~\bibnamefont {{Umeda}}},\ and\ \bibinfo {author} {\bibfnamefont {H.}~\bibnamefont {{Yanagisawa}}},\ }\bibfield  {title} {\bibinfo {title} {{JWST, ALMA, and Keck Spectroscopic Constraints on the UV Luminosity Functions at z {\ensuremath{\sim}} 7{\textendash}14: Clumpiness and Compactness of the Brightest Galaxies in the Early Universe}},\ }\href {https://doi.org/10.3847/1538-4357/ad9b2c} {\bibfield  {journal} {\bibinfo  {journal} {\apj}\ }\textbf {\bibinfo {volume} {980}},\ \bibinfo {eid} {138} (\bibinfo {year} {2025})},\ \Eprint {https://arxiv.org/abs/2406.18352} {arXiv:2406.18352 [astro-ph.GA]} \BibitemShut {NoStop}%
\bibitem [{\citenamefont {{Ferrarese}}\ and\ \citenamefont {{Merritt}}(2000)}]{Ferrarese2000}%
  \BibitemOpen
  \bibfield  {author} {\bibinfo {author} {\bibfnamefont {L.}~\bibnamefont {{Ferrarese}}}\ and\ \bibinfo {author} {\bibfnamefont {D.}~\bibnamefont {{Merritt}}},\ }\bibfield  {title} {\bibinfo {title} {{A Fundamental Relation between Supermassive Black Holes and Their Host Galaxies}},\ }\href {https://doi.org/10.1086/312838} {\bibfield  {journal} {\bibinfo  {journal} {\apjl}\ }\textbf {\bibinfo {volume} {539}},\ \bibinfo {pages} {L9} (\bibinfo {year} {2000})},\ \Eprint {https://arxiv.org/abs/astro-ph/0006053} {arXiv:astro-ph/0006053 [astro-ph]} \BibitemShut {NoStop}%
\bibitem [{\citenamefont {{Tremaine}}\ \emph {et~al.}(2002)\citenamefont {{Tremaine}}, \citenamefont {{Gebhardt}}, \citenamefont {{Bender}}, \citenamefont {{Bower}}, \citenamefont {{Dressler}}, \citenamefont {{Faber}}, \citenamefont {{Filippenko}}, \citenamefont {{Green}}, \citenamefont {{Grillmair}}, \citenamefont {{Ho}}, \citenamefont {{Kormendy}}, \citenamefont {{Lauer}}, \citenamefont {{Magorrian}}, \citenamefont {{Pinkney}},\ and\ \citenamefont {{Richstone}}}]{Tremaine2002}%
  \BibitemOpen
  \bibfield  {author} {\bibinfo {author} {\bibfnamefont {S.}~\bibnamefont {{Tremaine}}}, \bibinfo {author} {\bibfnamefont {K.}~\bibnamefont {{Gebhardt}}}, \bibinfo {author} {\bibfnamefont {R.}~\bibnamefont {{Bender}}}, \bibinfo {author} {\bibfnamefont {G.}~\bibnamefont {{Bower}}}, \bibinfo {author} {\bibfnamefont {A.}~\bibnamefont {{Dressler}}}, \bibinfo {author} {\bibfnamefont {S.~M.}\ \bibnamefont {{Faber}}}, \bibinfo {author} {\bibfnamefont {A.~V.}\ \bibnamefont {{Filippenko}}}, \bibinfo {author} {\bibfnamefont {R.}~\bibnamefont {{Green}}}, \bibinfo {author} {\bibfnamefont {C.}~\bibnamefont {{Grillmair}}}, \bibinfo {author} {\bibfnamefont {L.~C.}\ \bibnamefont {{Ho}}}, \bibinfo {author} {\bibfnamefont {J.}~\bibnamefont {{Kormendy}}}, \bibinfo {author} {\bibfnamefont {T.~R.}\ \bibnamefont {{Lauer}}}, \bibinfo {author} {\bibfnamefont {J.}~\bibnamefont {{Magorrian}}}, \bibinfo {author} {\bibfnamefont {J.}~\bibnamefont {{Pinkney}}},\ and\ \bibinfo {author} {\bibfnamefont {D.}~\bibnamefont {{Richstone}}},\
  }\bibfield  {title} {\bibinfo {title} {{The Slope of the Black Hole Mass versus Velocity Dispersion Correlation}},\ }\href {https://doi.org/10.1086/341002} {\bibfield  {journal} {\bibinfo  {journal} {\apj}\ }\textbf {\bibinfo {volume} {574}},\ \bibinfo {pages} {740} (\bibinfo {year} {2002})},\ \Eprint {https://arxiv.org/abs/astro-ph/0203468} {arXiv:astro-ph/0203468 [astro-ph]} \BibitemShut {NoStop}%
\bibitem [{\citenamefont {{Volonteri}}\ and\ \citenamefont {{Natarajan}}(2009)}]{VolNat2009}%
  \BibitemOpen
  \bibfield  {author} {\bibinfo {author} {\bibfnamefont {M.}~\bibnamefont {{Volonteri}}}\ and\ \bibinfo {author} {\bibfnamefont {P.}~\bibnamefont {{Natarajan}}},\ }\bibfield  {title} {\bibinfo {title} {{Journey to the M$_{BH} - \sigma$ relation: the fate of low-mass black holes in the Universe}},\ }\href {https://doi.org/10.1111/j.1365-2966.2009.15577.x} {\bibfield  {journal} {\bibinfo  {journal} {\mnras}\ }\textbf {\bibinfo {volume} {400}},\ \bibinfo {pages} {1911} (\bibinfo {year} {2009})},\ \Eprint {https://arxiv.org/abs/0903.2262} {arXiv:0903.2262 [astro-ph.CO]} \BibitemShut {NoStop}%
\bibitem [{\citenamefont {{Magorrian}}\ \emph {et~al.}(1998)\citenamefont {{Magorrian}}, \citenamefont {{Tremaine}}, \citenamefont {{Richstone}}, \citenamefont {{Bender}}, \citenamefont {{Bower}}, \citenamefont {{Dressler}}, \citenamefont {{Faber}}, \citenamefont {{Gebhardt}}, \citenamefont {{Green}}, \citenamefont {{Grillmair}}, \citenamefont {{Kormendy}},\ and\ \citenamefont {{Lauer}}}]{Magorrian1998}%
  \BibitemOpen
  \bibfield  {author} {\bibinfo {author} {\bibfnamefont {J.}~\bibnamefont {{Magorrian}}}, \bibinfo {author} {\bibfnamefont {S.}~\bibnamefont {{Tremaine}}}, \bibinfo {author} {\bibfnamefont {D.}~\bibnamefont {{Richstone}}}, \bibinfo {author} {\bibfnamefont {R.}~\bibnamefont {{Bender}}}, \bibinfo {author} {\bibfnamefont {G.}~\bibnamefont {{Bower}}}, \bibinfo {author} {\bibfnamefont {A.}~\bibnamefont {{Dressler}}}, \bibinfo {author} {\bibfnamefont {S.~M.}\ \bibnamefont {{Faber}}}, \bibinfo {author} {\bibfnamefont {K.}~\bibnamefont {{Gebhardt}}}, \bibinfo {author} {\bibfnamefont {R.}~\bibnamefont {{Green}}}, \bibinfo {author} {\bibfnamefont {C.}~\bibnamefont {{Grillmair}}}, \bibinfo {author} {\bibfnamefont {J.}~\bibnamefont {{Kormendy}}},\ and\ \bibinfo {author} {\bibfnamefont {T.}~\bibnamefont {{Lauer}}},\ }\bibfield  {title} {\bibinfo {title} {{The Demography of Massive Dark Objects in Galaxy Centers}},\ }\href {https://doi.org/10.1086/300353} {\bibfield  {journal} {\bibinfo  {journal} {Astronomical Journal}\
  }\textbf {\bibinfo {volume} {115}},\ \bibinfo {pages} {2285} (\bibinfo {year} {1998})},\ \Eprint {https://arxiv.org/abs/astro-ph/9708072} {arXiv:astro-ph/9708072 [astro-ph]} \BibitemShut {NoStop}%
\bibitem [{\citenamefont {{H{\"a}ring}}\ and\ \citenamefont {{Rix}}(2004)}]{Haring2004}%
  \BibitemOpen
  \bibfield  {author} {\bibinfo {author} {\bibfnamefont {N.}~\bibnamefont {{H{\"a}ring}}}\ and\ \bibinfo {author} {\bibfnamefont {H.-W.}\ \bibnamefont {{Rix}}},\ }\bibfield  {title} {\bibinfo {title} {{On the Black Hole Mass-Bulge Mass Relation}},\ }\href {https://doi.org/10.1086/383567} {\bibfield  {journal} {\bibinfo  {journal} {\apjl}\ }\textbf {\bibinfo {volume} {604}},\ \bibinfo {pages} {L89} (\bibinfo {year} {2004})},\ \Eprint {https://arxiv.org/abs/astro-ph/0402376} {arXiv:astro-ph/0402376 [astro-ph]} \BibitemShut {NoStop}%
\bibitem [{\citenamefont {{Ferrarese}}(2002)}]{Ferrarese2002}%
  \BibitemOpen
  \bibfield  {author} {\bibinfo {author} {\bibfnamefont {L.}~\bibnamefont {{Ferrarese}}},\ }\bibfield  {title} {\bibinfo {title} {{Beyond the Bulge: A Fundamental Relation between Supermassive Black Holes and Dark Matter Halos}},\ }\href {https://doi.org/10.1086/342308} {\bibfield  {journal} {\bibinfo  {journal} {\apj}\ }\textbf {\bibinfo {volume} {578}},\ \bibinfo {pages} {90} (\bibinfo {year} {2002})},\ \Eprint {https://arxiv.org/abs/astro-ph/0203469} {arXiv:astro-ph/0203469 [astro-ph]} \BibitemShut {NoStop}%
\bibitem [{\citenamefont {{Baes}}\ \emph {et~al.}(2003)\citenamefont {{Baes}}, \citenamefont {{Buyle}}, \citenamefont {{Hau}},\ and\ \citenamefont {{Dejonghe}}}]{Baes2003}%
  \BibitemOpen
  \bibfield  {author} {\bibinfo {author} {\bibfnamefont {M.}~\bibnamefont {{Baes}}}, \bibinfo {author} {\bibfnamefont {P.}~\bibnamefont {{Buyle}}}, \bibinfo {author} {\bibfnamefont {G.~K.~T.}\ \bibnamefont {{Hau}}},\ and\ \bibinfo {author} {\bibfnamefont {H.}~\bibnamefont {{Dejonghe}}},\ }\bibfield  {title} {\bibinfo {title} {{Observational evidence for a connection between supermassive black holes and dark matter haloes}},\ }\href {https://doi.org/10.1046/j.1365-8711.2003.06680.x} {\bibfield  {journal} {\bibinfo  {journal} {\mnras}\ }\textbf {\bibinfo {volume} {341}},\ \bibinfo {pages} {L44} (\bibinfo {year} {2003})},\ \Eprint {https://arxiv.org/abs/astro-ph/0303628} {arXiv:astro-ph/0303628 [astro-ph]} \BibitemShut {NoStop}%
\bibitem [{\citenamefont {{Filloux}}\ \emph {et~al.}(2010)\citenamefont {{Filloux}}, \citenamefont {{Durier}}, \citenamefont {{Pacheco}},\ and\ \citenamefont {{Silk}}}]{Silk2010}%
  \BibitemOpen
  \bibfield  {author} {\bibinfo {author} {\bibfnamefont {C.}~\bibnamefont {{Filloux}}}, \bibinfo {author} {\bibfnamefont {F.}~\bibnamefont {{Durier}}}, \bibinfo {author} {\bibfnamefont {J.~A.~F.}\ \bibnamefont {{Pacheco}}},\ and\ \bibinfo {author} {\bibfnamefont {J.}~\bibnamefont {{Silk}}},\ }\bibfield  {title} {\bibinfo {title} {{Evolution of Supermassive Black Holes from Cosmological Simulations}},\ }\href {https://doi.org/10.1142/S0218271810017603} {\bibfield  {journal} {\bibinfo  {journal} {International Journal of Modern Physics D}\ }\textbf {\bibinfo {volume} {19}},\ \bibinfo {pages} {1233} (\bibinfo {year} {2010})},\ \Eprint {https://arxiv.org/abs/0912.2223} {arXiv:0912.2223 [astro-ph.CO]} \BibitemShut {NoStop}%
\bibitem [{\citenamefont {{Volonteri}}\ \emph {et~al.}(2009)\citenamefont {{Volonteri}}, \citenamefont {{Miller}},\ and\ \citenamefont {{Dotti}}}]{Volonteri2009}%
  \BibitemOpen
  \bibfield  {author} {\bibinfo {author} {\bibfnamefont {M.}~\bibnamefont {{Volonteri}}}, \bibinfo {author} {\bibfnamefont {J.~M.}\ \bibnamefont {{Miller}}},\ and\ \bibinfo {author} {\bibfnamefont {M.}~\bibnamefont {{Dotti}}},\ }\bibfield  {title} {\bibinfo {title} {{Sub-Parsec Supermassive Binary Quasars: Expectations at $z < 1$}},\ }\href {https://doi.org/10.1088/0004-637X/703/1/L86} {\bibfield  {journal} {\bibinfo  {journal} {\apjl}\ }\textbf {\bibinfo {volume} {703}},\ \bibinfo {pages} {L86} (\bibinfo {year} {2009})},\ \Eprint {https://arxiv.org/abs/0903.3947} {arXiv:0903.3947 [astro-ph.CO]} \BibitemShut {NoStop}%
\bibitem [{\citenamefont {{Krolik}}\ \emph {et~al.}(2019)\citenamefont {{Krolik}}, \citenamefont {{Volonteri}}, \citenamefont {{Dubois}},\ and\ \citenamefont {{Devriendt}}}]{Krolik2019}%
  \BibitemOpen
  \bibfield  {author} {\bibinfo {author} {\bibfnamefont {J.~H.}\ \bibnamefont {{Krolik}}}, \bibinfo {author} {\bibfnamefont {M.}~\bibnamefont {{Volonteri}}}, \bibinfo {author} {\bibfnamefont {Y.}~\bibnamefont {{Dubois}}},\ and\ \bibinfo {author} {\bibfnamefont {J.}~\bibnamefont {{Devriendt}}},\ }\bibfield  {title} {\bibinfo {title} {{Population Estimates for Electromagnetically Distinguishable Supermassive Binary Black Holes}},\ }\href {https://doi.org/10.3847/1538-4357/ab24c9} {\bibfield  {journal} {\bibinfo  {journal} {\apj}\ }\textbf {\bibinfo {volume} {879}},\ \bibinfo {eid} {110} (\bibinfo {year} {2019})},\ \Eprint {https://arxiv.org/abs/1905.10450} {arXiv:1905.10450 [astro-ph.GA]} \BibitemShut {NoStop}%
\bibitem [{\citenamefont {{De Rosa}}\ \emph {et~al.}(2019)\citenamefont {{De Rosa}}, \citenamefont {{Vignali}}, \citenamefont {{Bogdanovi{\'c}}}, \citenamefont {{Capelo}}, \citenamefont {{Charisi}}, \citenamefont {{Dotti}}, \citenamefont {{Husemann}}, \citenamefont {{Lusso}}, \citenamefont {{Mayer}}, \citenamefont {{Paragi}}, \citenamefont {{Runnoe}}, \citenamefont {{Sesana}}, \citenamefont {{Steinborn}}, \citenamefont {{Bianchi}}, \citenamefont {{Colpi}}, \citenamefont {{del Valle}}, \citenamefont {{Frey}}, \citenamefont {{Gab{\'a}nyi}}, \citenamefont {{Giustini}}, \citenamefont {{Guainazzi}}, \citenamefont {{Haiman}}, \citenamefont {{Herrera Ruiz}}, \citenamefont {{Herrero-Illana}}, \citenamefont {{Iwasawa}}, \citenamefont {{Komossa}}, \citenamefont {{Lena}}, \citenamefont {{Loiseau}}, \citenamefont {{Perez-Torres}}, \citenamefont {{Piconcelli}},\ and\ \citenamefont {{Volonteri}}}]{DeRosa2019}%
  \BibitemOpen
  \bibfield  {author} {\bibinfo {author} {\bibfnamefont {A.}~\bibnamefont {{De Rosa}}}, \bibinfo {author} {\bibfnamefont {C.}~\bibnamefont {{Vignali}}}, \bibinfo {author} {\bibfnamefont {T.}~\bibnamefont {{Bogdanovi{\'c}}}}, \bibinfo {author} {\bibfnamefont {P.~R.}\ \bibnamefont {{Capelo}}}, \bibinfo {author} {\bibfnamefont {M.}~\bibnamefont {{Charisi}}}, \bibinfo {author} {\bibfnamefont {M.}~\bibnamefont {{Dotti}}}, \bibinfo {author} {\bibfnamefont {B.}~\bibnamefont {{Husemann}}}, \bibinfo {author} {\bibfnamefont {E.}~\bibnamefont {{Lusso}}}, \bibinfo {author} {\bibfnamefont {L.}~\bibnamefont {{Mayer}}}, \bibinfo {author} {\bibfnamefont {Z.}~\bibnamefont {{Paragi}}}, \bibinfo {author} {\bibfnamefont {J.}~\bibnamefont {{Runnoe}}}, \bibinfo {author} {\bibfnamefont {A.}~\bibnamefont {{Sesana}}}, \bibinfo {author} {\bibfnamefont {L.}~\bibnamefont {{Steinborn}}}, \bibinfo {author} {\bibfnamefont {S.}~\bibnamefont {{Bianchi}}}, \bibinfo {author} {\bibfnamefont {M.}~\bibnamefont {{Colpi}}}, \bibinfo {author}
  {\bibfnamefont {L.}~\bibnamefont {{del Valle}}}, \bibinfo {author} {\bibfnamefont {S.}~\bibnamefont {{Frey}}}, \bibinfo {author} {\bibfnamefont {K.~{\'E}.}\ \bibnamefont {{Gab{\'a}nyi}}}, \bibinfo {author} {\bibfnamefont {M.}~\bibnamefont {{Giustini}}}, \bibinfo {author} {\bibfnamefont {M.}~\bibnamefont {{Guainazzi}}}, \bibinfo {author} {\bibfnamefont {Z.}~\bibnamefont {{Haiman}}}, \bibinfo {author} {\bibfnamefont {N.}~\bibnamefont {{Herrera Ruiz}}}, \bibinfo {author} {\bibfnamefont {R.}~\bibnamefont {{Herrero-Illana}}}, \bibinfo {author} {\bibfnamefont {K.}~\bibnamefont {{Iwasawa}}}, \bibinfo {author} {\bibfnamefont {S.}~\bibnamefont {{Komossa}}}, \bibinfo {author} {\bibfnamefont {D.}~\bibnamefont {{Lena}}}, \bibinfo {author} {\bibfnamefont {N.}~\bibnamefont {{Loiseau}}}, \bibinfo {author} {\bibfnamefont {M.}~\bibnamefont {{Perez-Torres}}}, \bibinfo {author} {\bibfnamefont {E.}~\bibnamefont {{Piconcelli}}},\ and\ \bibinfo {author} {\bibfnamefont {M.}~\bibnamefont {{Volonteri}}},\ }\bibfield  {title}
  {\bibinfo {title} {{The quest for dual and binary supermassive black holes: A multi-messenger view}},\ }\href {https://doi.org/10.1016/j.newar.2020.101525} {\bibfield  {journal} {\bibinfo  {journal} {New Astronomy Reviews}\ }\textbf {\bibinfo {volume} {86}},\ \bibinfo {eid} {101525} (\bibinfo {year} {2019})},\ \Eprint {https://arxiv.org/abs/2001.06293} {arXiv:2001.06293 [astro-ph.GA]} \BibitemShut {NoStop}%
\bibitem [{\citenamefont {{Ni}}\ \emph {et~al.}(2020)\citenamefont {{Ni}}, \citenamefont {{Di Matteo}}, \citenamefont {{Gilli}}, \citenamefont {{Croft}}, \citenamefont {{Feng}},\ and\ \citenamefont {{Norman}}}]{Ni20}%
  \BibitemOpen
  \bibfield  {author} {\bibinfo {author} {\bibfnamefont {Y.}~\bibnamefont {{Ni}}}, \bibinfo {author} {\bibfnamefont {T.}~\bibnamefont {{Di Matteo}}}, \bibinfo {author} {\bibfnamefont {R.}~\bibnamefont {{Gilli}}}, \bibinfo {author} {\bibfnamefont {R.~A.~C.}\ \bibnamefont {{Croft}}}, \bibinfo {author} {\bibfnamefont {Y.}~\bibnamefont {{Feng}}},\ and\ \bibinfo {author} {\bibfnamefont {C.}~\bibnamefont {{Norman}}},\ }\bibfield  {title} {\bibinfo {title} {{QSO obscuration at high redshift (z {\ensuremath{\gtrsim}} 7): predictions from the BLUETIDES simulation}},\ }\href {https://doi.org/10.1093/mnras/staa1313} {\bibfield  {journal} {\bibinfo  {journal} {\mnras}\ }\textbf {\bibinfo {volume} {495}},\ \bibinfo {pages} {2135} (\bibinfo {year} {2020})},\ \Eprint {https://arxiv.org/abs/1912.03780} {arXiv:1912.03780 [astro-ph.GA]} \BibitemShut {NoStop}%
\bibitem [{\citenamefont {{Kardashev}}\ \emph {et~al.}(2014)\citenamefont {{Kardashev}}, \citenamefont {{Novikov}}, \citenamefont {{Lukash}}, \citenamefont {{Pilipenko}}, \citenamefont {{Mikheeva}}, \citenamefont {{Bisikalo}}, \citenamefont {{Wiebe}}, \citenamefont {{Doroshkevich}}, \citenamefont {{Zasov}}, \citenamefont {{Zinchenko}}, \citenamefont {{Ivanov}}, \citenamefont {{Kostenko}}, \citenamefont {{Larchenkova}}, \citenamefont {{Likhachev}}, \citenamefont {{Malov}}, \citenamefont {{Malofeev}}, \citenamefont {{Pozanenko}}, \citenamefont {{Smirnov}}, \citenamefont {{Sobolev}}, \citenamefont {{Cherepashchuk}},\ and\ \citenamefont {{Shchekinov}}}]{Millimetron14}%
  \BibitemOpen
  \bibfield  {author} {\bibinfo {author} {\bibfnamefont {N.~S.}\ \bibnamefont {{Kardashev}}}, \bibinfo {author} {\bibfnamefont {I.~D.}\ \bibnamefont {{Novikov}}}, \bibinfo {author} {\bibfnamefont {V.~N.}\ \bibnamefont {{Lukash}}}, \bibinfo {author} {\bibfnamefont {S.~V.}\ \bibnamefont {{Pilipenko}}}, \bibinfo {author} {\bibfnamefont {E.~V.}\ \bibnamefont {{Mikheeva}}}, \bibinfo {author} {\bibfnamefont {D.~V.}\ \bibnamefont {{Bisikalo}}}, \bibinfo {author} {\bibfnamefont {D.~S.}\ \bibnamefont {{Wiebe}}}, \bibinfo {author} {\bibfnamefont {A.~G.}\ \bibnamefont {{Doroshkevich}}}, \bibinfo {author} {\bibfnamefont {A.~V.}\ \bibnamefont {{Zasov}}}, \bibinfo {author} {\bibfnamefont {I.~I.}\ \bibnamefont {{Zinchenko}}}, \bibinfo {author} {\bibfnamefont {P.~B.}\ \bibnamefont {{Ivanov}}}, \bibinfo {author} {\bibfnamefont {V.~I.}\ \bibnamefont {{Kostenko}}}, \bibinfo {author} {\bibfnamefont {T.~I.}\ \bibnamefont {{Larchenkova}}}, \bibinfo {author} {\bibfnamefont {S.~F.}\ \bibnamefont {{Likhachev}}}, \bibinfo {author}
  {\bibfnamefont {I.~F.}\ \bibnamefont {{Malov}}}, \bibinfo {author} {\bibfnamefont {V.~M.}\ \bibnamefont {{Malofeev}}}, \bibinfo {author} {\bibfnamefont {A.~S.}\ \bibnamefont {{Pozanenko}}}, \bibinfo {author} {\bibfnamefont {A.~V.}\ \bibnamefont {{Smirnov}}}, \bibinfo {author} {\bibfnamefont {A.~M.}\ \bibnamefont {{Sobolev}}}, \bibinfo {author} {\bibfnamefont {A.~M.}\ \bibnamefont {{Cherepashchuk}}},\ and\ \bibinfo {author} {\bibfnamefont {Y.~A.}\ \bibnamefont {{Shchekinov}}},\ }\bibfield  {title} {\bibinfo {title} {{Review of scientific topics for the Millimetron space observatory}},\ }\href {https://doi.org/10.3367/UFNe.0184.201412c.1319} {\bibfield  {journal} {\bibinfo  {journal} {Physics Uspekhi}\ }\textbf {\bibinfo {volume} {57}},\ \bibinfo {eid} {1199-1228} (\bibinfo {year} {2014})},\ \Eprint {https://arxiv.org/abs/1502.06071} {arXiv:1502.06071 [astro-ph.IM]} \BibitemShut {NoStop}%
\bibitem [{\citenamefont {{Novikov}}\ \emph {et~al.}(2021)\citenamefont {{Novikov}}, \citenamefont {{Likhachev}}, \citenamefont {{Shchekinov}}, \citenamefont {{Andrianov}}, \citenamefont {{Baryshev}}, \citenamefont {{Vasyunin}}, \citenamefont {{Wiebe}}, \citenamefont {{Graauw}}, \citenamefont {{Doroshkevich}}, \citenamefont {{Zinchenko}}, \citenamefont {{Kardashev}}, \citenamefont {{Kostenko}}, \citenamefont {{Larchenkova}}, \citenamefont {{Likhacheva}}, \citenamefont {{Lyakhovets}}, \citenamefont {{Novikov}}, \citenamefont {{Pilipenko}}, \citenamefont {{Punanova}}, \citenamefont {{Rudnitsky}}, \citenamefont {{Smirnov}},\ and\ \citenamefont {{Shematovich}}}]{Millimetron21}%
  \BibitemOpen
  \bibfield  {author} {\bibinfo {author} {\bibfnamefont {I.~D.}\ \bibnamefont {{Novikov}}}, \bibinfo {author} {\bibfnamefont {S.~F.}\ \bibnamefont {{Likhachev}}}, \bibinfo {author} {\bibfnamefont {Y.~A.}\ \bibnamefont {{Shchekinov}}}, \bibinfo {author} {\bibfnamefont {A.~S.}\ \bibnamefont {{Andrianov}}}, \bibinfo {author} {\bibfnamefont {A.~M.}\ \bibnamefont {{Baryshev}}}, \bibinfo {author} {\bibfnamefont {A.~I.}\ \bibnamefont {{Vasyunin}}}, \bibinfo {author} {\bibfnamefont {D.~Z.}\ \bibnamefont {{Wiebe}}}, \bibinfo {author} {\bibfnamefont {T.~d.}\ \bibnamefont {{Graauw}}}, \bibinfo {author} {\bibfnamefont {A.~G.}\ \bibnamefont {{Doroshkevich}}}, \bibinfo {author} {\bibfnamefont {I.~I.}\ \bibnamefont {{Zinchenko}}}, \bibinfo {author} {\bibfnamefont {N.~S.}\ \bibnamefont {{Kardashev}}}, \bibinfo {author} {\bibfnamefont {V.~I.}\ \bibnamefont {{Kostenko}}}, \bibinfo {author} {\bibfnamefont {T.~I.}\ \bibnamefont {{Larchenkova}}}, \bibinfo {author} {\bibfnamefont {L.~N.}\ \bibnamefont {{Likhacheva}}}, \bibinfo
  {author} {\bibfnamefont {A.~O.}\ \bibnamefont {{Lyakhovets}}}, \bibinfo {author} {\bibfnamefont {D.~I.}\ \bibnamefont {{Novikov}}}, \bibinfo {author} {\bibfnamefont {S.~V.}\ \bibnamefont {{Pilipenko}}}, \bibinfo {author} {\bibfnamefont {A.~F.}\ \bibnamefont {{Punanova}}}, \bibinfo {author} {\bibfnamefont {A.~G.}\ \bibnamefont {{Rudnitsky}}}, \bibinfo {author} {\bibfnamefont {A.~V.}\ \bibnamefont {{Smirnov}}},\ and\ \bibinfo {author} {\bibfnamefont {V.~I.}\ \bibnamefont {{Shematovich}}},\ }\bibfield  {title} {\bibinfo {title} {{Objectives of the Millimetron Space Observatory science program and technical capabilities of its realization}},\ }\href {https://doi.org/10.3367/UFNe.2020.12.038898} {\bibfield  {journal} {\bibinfo  {journal} {Physics Uspekhi}\ }\textbf {\bibinfo {volume} {64}},\ \bibinfo {pages} {386} (\bibinfo {year} {2021})}\BibitemShut {NoStop}%
\bibitem [{\citenamefont {{Press}}\ and\ \citenamefont {{Schechter}}(1974)}]{press}%
  \BibitemOpen
  \bibfield  {author} {\bibinfo {author} {\bibfnamefont {W.~H.}\ \bibnamefont {{Press}}}\ and\ \bibinfo {author} {\bibfnamefont {P.}~\bibnamefont {{Schechter}}},\ }\bibfield  {title} {\bibinfo {title} {{Formation of Galaxies and Clusters of Galaxies by Self-Similar Gravitational Condensation}},\ }\href {https://doi.org/10.1086/152650} {\bibfield  {journal} {\bibinfo  {journal} {\apj}\ }\textbf {\bibinfo {volume} {187}},\ \bibinfo {pages} {425} (\bibinfo {year} {1974})}\BibitemShut {NoStop}%
\bibitem [{\citenamefont {{Springel}}(2005)}]{gadget}%
  \BibitemOpen
  \bibfield  {author} {\bibinfo {author} {\bibfnamefont {V.}~\bibnamefont {{Springel}}},\ }\bibfield  {title} {\bibinfo {title} {{The cosmological simulation code GADGET-2}},\ }\href {https://doi.org/10.1111/j.1365-2966.2005.09655.x} {\bibfield  {journal} {\bibinfo  {journal} {\mnras}\ }\textbf {\bibinfo {volume} {364}},\ \bibinfo {pages} {1105} (\bibinfo {year} {2005})},\ \Eprint {https://arxiv.org/abs/astro-ph/0505010} {astro-ph/0505010} \BibitemShut {NoStop}%
\bibitem [{Note1()}]{Note1}%
  \BibitemOpen
  \bibinfo {note} {Https://github.com/ginnungagapgroup/ginnungagap}\BibitemShut {NoStop}%
\bibitem [{\citenamefont {Blas}\ \emph {et~al.}(2011)\citenamefont {Blas}, \citenamefont {Lesgourgues},\ and\ \citenamefont {Tram}}]{CLASS}%
  \BibitemOpen
  \bibfield  {author} {\bibinfo {author} {\bibfnamefont {D.}~\bibnamefont {Blas}}, \bibinfo {author} {\bibfnamefont {J.}~\bibnamefont {Lesgourgues}},\ and\ \bibinfo {author} {\bibfnamefont {T.}~\bibnamefont {Tram}},\ }\bibfield  {title} {\bibinfo {title} {The cosmic linear anisotropy solving system ({CLASS}). part {II}: Approximation schemes},\ }\href {https://doi.org/10.1088/1475-7516/2011/07/034} {\bibfield  {journal} {\bibinfo  {journal} {\jcap}\ }\textbf {\bibinfo {volume} {2011}},\ \bibinfo {pages} {034} (\bibinfo {year} {2011})}\BibitemShut {NoStop}%
\bibitem [{\citenamefont {{Knollmann}}\ and\ \citenamefont {{Knebe}}(2009)}]{AHF}%
  \BibitemOpen
  \bibfield  {author} {\bibinfo {author} {\bibfnamefont {S.~R.}\ \bibnamefont {{Knollmann}}}\ and\ \bibinfo {author} {\bibfnamefont {A.}~\bibnamefont {{Knebe}}},\ }\bibfield  {title} {\bibinfo {title} {{AHF: Amiga's Halo Finder}},\ }\href {https://doi.org/10.1088/0067-0049/182/2/608} {\bibfield  {journal} {\bibinfo  {journal} {\apjs}\ }\textbf {\bibinfo {volume} {182}},\ \bibinfo {pages} {608} (\bibinfo {year} {2009})},\ \Eprint {https://arxiv.org/abs/0904.3662} {arXiv:0904.3662 [astro-ph.CO]} \BibitemShut {NoStop}%
\bibitem [{\citenamefont {{Planck Collaboration}}\ \emph {et~al.}(2014)\citenamefont {{Planck Collaboration}}, \citenamefont {{Ade}}, \citenamefont {{Aghanim}}, \citenamefont {{Armitage-Caplan}}, \citenamefont {{Arnaud}}, \citenamefont {{Ashdown}}, \citenamefont {{Atrio-Barandela}}, \citenamefont {{Aumont}}, \citenamefont {{Baccigalupi}}, \citenamefont {{Banday}},\ and\ \citenamefont {et~al.}}]{planck}%
  \BibitemOpen
  \bibfield  {author} {\bibinfo {author} {\bibnamefont {{Planck Collaboration}}}, \bibinfo {author} {\bibfnamefont {P.~A.~R.}\ \bibnamefont {{Ade}}}, \bibinfo {author} {\bibfnamefont {N.}~\bibnamefont {{Aghanim}}}, \bibinfo {author} {\bibfnamefont {C.}~\bibnamefont {{Armitage-Caplan}}}, \bibinfo {author} {\bibfnamefont {M.}~\bibnamefont {{Arnaud}}}, \bibinfo {author} {\bibfnamefont {M.}~\bibnamefont {{Ashdown}}}, \bibinfo {author} {\bibfnamefont {F.}~\bibnamefont {{Atrio-Barandela}}}, \bibinfo {author} {\bibfnamefont {J.}~\bibnamefont {{Aumont}}}, \bibinfo {author} {\bibfnamefont {C.}~\bibnamefont {{Baccigalupi}}}, \bibinfo {author} {\bibfnamefont {A.~J.}\ \bibnamefont {{Banday}}},\ and\ \bibinfo {author} {\bibnamefont {et~al.}},\ }\bibfield  {title} {\bibinfo {title} {{Planck 2013 results. XVI. Cosmological parameters}},\ }\href {https://doi.org/10.1051/0004-6361/201321591} {\bibfield  {journal} {\bibinfo  {journal} {\aap}\ }\textbf {\bibinfo {volume} {571}},\ \bibinfo {eid} {A16} (\bibinfo {year} {2014})},\
  \Eprint {https://arxiv.org/abs/1303.5076} {arXiv:1303.5076 [astro-ph.CO]} \BibitemShut {NoStop}%
\bibitem [{\citenamefont {Zhang}\ \emph {et~al.}(2022)\citenamefont {Zhang}, \citenamefont {Behroozi}, \citenamefont {Volonteri}, \citenamefont {Silk}, \citenamefont {Fan}, \citenamefont {Hopkins}, \citenamefont {Yang},\ and\ \citenamefont {Aird}}]{trinity_i}%
  \BibitemOpen
  \bibfield  {author} {\bibinfo {author} {\bibfnamefont {H.}~\bibnamefont {Zhang}}, \bibinfo {author} {\bibfnamefont {P.}~\bibnamefont {Behroozi}}, \bibinfo {author} {\bibfnamefont {M.}~\bibnamefont {Volonteri}}, \bibinfo {author} {\bibfnamefont {J.}~\bibnamefont {Silk}}, \bibinfo {author} {\bibfnamefont {X.}~\bibnamefont {Fan}}, \bibinfo {author} {\bibfnamefont {P.~F.}\ \bibnamefont {Hopkins}}, \bibinfo {author} {\bibfnamefont {J.}~\bibnamefont {Yang}},\ and\ \bibinfo {author} {\bibfnamefont {J.}~\bibnamefont {Aird}},\ }\bibfield  {title} {\bibinfo {title} {Trinity i: self-consistently modelling the dark matter halo–galaxy–supermassive black hole connection from $z = 0–10$},\ }\href {https://doi.org/10.1093/mnras/stac2633} {\bibfield  {journal} {\bibinfo  {journal} {\mnras}\ }\textbf {\bibinfo {volume} {518}},\ \bibinfo {pages} {2123} (\bibinfo {year} {2022})}\BibitemShut {NoStop}%
\bibitem [{\citenamefont {Zhang}\ \emph {et~al.}(2023)\citenamefont {Zhang}, \citenamefont {Behroozi}, \citenamefont {Volonteri}, \citenamefont {Silk}, \citenamefont {Fan}, \citenamefont {Aird}, \citenamefont {Yang},\ and\ \citenamefont {Hopkins}}]{trinity_ii}%
  \BibitemOpen
  \bibfield  {author} {\bibinfo {author} {\bibfnamefont {H.}~\bibnamefont {Zhang}}, \bibinfo {author} {\bibfnamefont {P.}~\bibnamefont {Behroozi}}, \bibinfo {author} {\bibfnamefont {M.}~\bibnamefont {Volonteri}}, \bibinfo {author} {\bibfnamefont {J.}~\bibnamefont {Silk}}, \bibinfo {author} {\bibfnamefont {X.}~\bibnamefont {Fan}}, \bibinfo {author} {\bibfnamefont {J.}~\bibnamefont {Aird}}, \bibinfo {author} {\bibfnamefont {J.}~\bibnamefont {Yang}},\ and\ \bibinfo {author} {\bibfnamefont {P.~F.}\ \bibnamefont {Hopkins}},\ }\bibfield  {title} {\bibinfo {title} {Trinity ii: The luminosity-dependent bias of the supermassive black hole mass–galaxy mass relation for bright quasars at $z = 6$},\ }\href@noop {} {\bibfield  {journal} {\bibinfo  {journal} {\mnras\, Lett.}\ }\textbf {\bibinfo {volume} {523}},\ \bibinfo {pages} {L69} (\bibinfo {year} {2023})}\BibitemShut {NoStop}%
\bibitem [{\citenamefont {{Zhang}}\ \emph {et~al.}(2024)\citenamefont {{Zhang}}, \citenamefont {{Behroozi}}, \citenamefont {{Volonteri}}, \citenamefont {{Silk}}, \citenamefont {{Fan}}, \citenamefont {{Aird}}, \citenamefont {{Yang}},\ and\ \citenamefont {{Hopkins}}}]{trinity_iii}%
  \BibitemOpen
  \bibfield  {author} {\bibinfo {author} {\bibfnamefont {H.}~\bibnamefont {{Zhang}}}, \bibinfo {author} {\bibfnamefont {P.}~\bibnamefont {{Behroozi}}}, \bibinfo {author} {\bibfnamefont {M.}~\bibnamefont {{Volonteri}}}, \bibinfo {author} {\bibfnamefont {J.}~\bibnamefont {{Silk}}}, \bibinfo {author} {\bibfnamefont {X.}~\bibnamefont {{Fan}}}, \bibinfo {author} {\bibfnamefont {J.}~\bibnamefont {{Aird}}}, \bibinfo {author} {\bibfnamefont {J.}~\bibnamefont {{Yang}}},\ and\ \bibinfo {author} {\bibfnamefont {P.~F.}\ \bibnamefont {{Hopkins}}},\ }\bibfield  {title} {\bibinfo {title} {{TRINITY - III. Quasar luminosity functions decomposed by halo, galaxy, and black hole masses as well as Eddington ratios from $z = 0-10$}},\ }\href {https://doi.org/10.1093/mnras/stae655} {\bibfield  {journal} {\bibinfo  {journal} {\mnras}\ }\textbf {\bibinfo {volume} {529}},\ \bibinfo {pages} {2777} (\bibinfo {year} {2024})},\ \Eprint {https://arxiv.org/abs/2305.19315} {arXiv:2305.19315 [astro-ph.GA]} \BibitemShut {NoStop}%
\bibitem [{\citenamefont {{Zhang}}\ \emph {et~al.}(2025)\citenamefont {{Zhang}}, \citenamefont {{Behroozi}}, \citenamefont {{Volonteri}}, \citenamefont {{Silk}}, \citenamefont {{Fan}}, \citenamefont {{Aird}}, \citenamefont {{Yang}}, \citenamefont {{Wang}},\ and\ \citenamefont {{Hopkins}}}]{trinity_iv}%
  \BibitemOpen
  \bibfield  {author} {\bibinfo {author} {\bibfnamefont {H.}~\bibnamefont {{Zhang}}}, \bibinfo {author} {\bibfnamefont {P.}~\bibnamefont {{Behroozi}}}, \bibinfo {author} {\bibfnamefont {M.}~\bibnamefont {{Volonteri}}}, \bibinfo {author} {\bibfnamefont {J.}~\bibnamefont {{Silk}}}, \bibinfo {author} {\bibfnamefont {X.}~\bibnamefont {{Fan}}}, \bibinfo {author} {\bibfnamefont {J.}~\bibnamefont {{Aird}}}, \bibinfo {author} {\bibfnamefont {J.}~\bibnamefont {{Yang}}}, \bibinfo {author} {\bibfnamefont {F.}~\bibnamefont {{Wang}}},\ and\ \bibinfo {author} {\bibfnamefont {P.~F.}\ \bibnamefont {{Hopkins}}},\ }\bibfield  {title} {\bibinfo {title} {{TRINITY VI: connection between galaxy star formation rates and supermassive black hole accretion rates from $z = 0 - 10$}},\ }\href {https://doi.org/10.1093/mnras/staf260} {\bibfield  {journal} {\bibinfo  {journal} {\mnras}\ }\textbf {\bibinfo {volume} {538}},\ \bibinfo {pages} {503} (\bibinfo {year} {2025})},\ \Eprint {https://arxiv.org/abs/2409.16347} {arXiv:2409.16347
  [astro-ph.GA]} \BibitemShut {NoStop}%
\bibitem [{Note2()}]{Note2}%
  \BibitemOpen
  \bibinfo {note} {The \protect \texttt {TRINITY} code is publicly available at \protect \texttt {https://github.com/HaowenZhang/TRINITY}.}\BibitemShut {Stop}%
\bibitem [{\citenamefont {{Sheth}}\ and\ \citenamefont {{Tormen}}(1999)}]{Sheth99}%
  \BibitemOpen
  \bibfield  {author} {\bibinfo {author} {\bibfnamefont {R.~K.}\ \bibnamefont {{Sheth}}}\ and\ \bibinfo {author} {\bibfnamefont {G.}~\bibnamefont {{Tormen}}},\ }\bibfield  {title} {\bibinfo {title} {{Large-scale bias and the peak background split}},\ }\href {https://doi.org/10.1046/j.1365-8711.1999.02692.x} {\bibfield  {journal} {\bibinfo  {journal} {\mnras}\ }\textbf {\bibinfo {volume} {308}},\ \bibinfo {pages} {119} (\bibinfo {year} {1999})},\ \Eprint {https://arxiv.org/abs/astro-ph/9901122} {arXiv:astro-ph/9901122 [astro-ph]} \BibitemShut {NoStop}%
\bibitem [{\citenamefont {Maiolino}\ \emph {et~al.}(2024)\citenamefont {Maiolino}, \citenamefont {Scholtz}, \citenamefont {Witstok},\ and\ \citenamefont {et~al.}}]{maiolino2024}%
  \BibitemOpen
  \bibfield  {author} {\bibinfo {author} {\bibfnamefont {R.}~\bibnamefont {Maiolino}}, \bibinfo {author} {\bibfnamefont {J.}~\bibnamefont {Scholtz}}, \bibinfo {author} {\bibfnamefont {J.}~\bibnamefont {Witstok}},\ and\ \bibinfo {author} {\bibnamefont {et~al.}},\ }\bibfield  {title} {\bibinfo {title} {A small and vigorous black hole in the early universe},\ }\href@noop {} {\bibfield  {journal} {\bibinfo  {journal} {Nature}\ }\textbf {\bibinfo {volume} {627}},\ \bibinfo {pages} {59} (\bibinfo {year} {2024})}\BibitemShut {NoStop}%
\bibitem [{\citenamefont {Benson}(2020)}]{benson2020}%
  \BibitemOpen
  \bibfield  {author} {\bibinfo {author} {\bibfnamefont {A.~J.}\ \bibnamefont {Benson}},\ }\bibfield  {title} {\bibinfo {title} {The normalization and slope of the dark matter (sub-)halo mass function on sub-galactic scales},\ }\href {https://doi.org/10.1093/mnras/staa341} {\bibfield  {journal} {\bibinfo  {journal} {Monthly Notices of the Royal Astronomical Society}\ }\textbf {\bibinfo {volume} {493}},\ \bibinfo {pages} {1268–1276} (\bibinfo {year} {2020})}\BibitemShut {NoStop}%
\end{thebibliography}%

\end{document}